\documentclass[galaxies,article,accept,moreauthors]{Definitions/mdpi} 
\newcommand{\mnras}{MNRAS}

\newcommand{\nat}{Nature}

\firstpage{1} 
\pubvolume{1}
\issuenum{1}
\articlenumber{0}
\pubyear{2024}
\copyrightyear{2024}
\externaleditor{Academic Editor: Jose L. Gómez}
\datereceived{6 February 2024} 
\daterevised{5 June 2024} 
\dateaccepted{} 
\datepublished{} 
\hreflink{https://doi.org/} 

\Title{{
A Supermassive Binary Black Hole Candidate in Mrk~501
}}

\TitleCitation{A Supermassive Binary Black Hole Candidate in Mrk~501}

\Author{Gustavo 
 Magallanes-Guij\'on
 $^{\dagger}$\orcidA{} and 
  Sergio Mendoza $^{*,\dagger}$\orcidB{} }

\AuthorNames{Gustavo Magallanes-Guij\'on and Sergio Mendoza}

\AuthorCitation{Magallanes
-Guij\'on, G.; Mendoza, S.}

\address[1]{Instituto 
 de Astronom\'{\i}a, Universidad Nacional Aut\'onoma de M\'exico, 
AP~70-263, Ciudad Universitaria, \mbox{Ciudad de M\'exico} 
04510, Mexico 
\\
}

\corres{Correspondence: gustavo.magallanes.guijon@ciencias.unam.mx (G.M.-G.);
sergio@astro.unam.mx (S.M.)}

\firstnote{These authors 
 contributed equally to this work.}

\abstract{
Using multifrequency observations, from radio to \( \gamma \)$-$rays 
of the blazar Mrk~501, we constructed their corresponding
light curves and built periodograms using RobPer and Lomb--Scargle
algorithms.  Long-term variability was also
studied using the power density spectrum and the detrended function
analysis. Using the {software VARTOOLS},
 we also computed the analysis of
variance, box-least squares and discrete fourier transform. The result
of these techniques showed an achromatic periodicity \( \lesssim\)\(229
\, \text{d} \).  This, combined with the result of pink-color noise
in the spectra, led us to propose that the periodicity was produced
via a secondary eclipsing supermassive binary black hole orbiting the
primary one locked inside the central engine of Mrk~501.  We built a
relativistic eclipsing model of this phenomenon using Jacobi elliptical
functions, finding a periodic relativistic eclipse occurring every \(\sim\)\(224 \,\text{d} \) in all the studied wavebands. 
This implies that the frequency of the emitted gravitational waves
falls slightly above \(0.1 \)~mHz, well within the operational range of the
upcoming LISA space-based interferometer, and as such, these gravitational
waves must be considered as a prime science target for future LISA 
observations.
}


\begin{document}

\section{Introduction}
\label{introduction}

Among active galactic nuclei (AGN), blazars are objects that emit
variable non-thermal radiation throughout the electromagnetic~(EM)
spectrum~\citep{Padovani2012} with their jets pointing at an
angle of no more than $\sim$$30^{\circ}$ from the observer's line of
sight~\citep{Ulrich1997}. These extragalactic sources present total
luminosities in the range  of $10^{41}$--$10^{47}\, \text{ergs}/\text{s}
$ (see, e.g., 
 \cite{Blandford2019}).

The light curve variabilities in blazars are commonly classified as follows:
(a) intraday variability (IDV), corresponding to periods of over a day
or less~\citep{Wagner1995}---they are also called intra-night variability
or micro-variability~\citep{Miller1989Natur}; (b) short-term variability
(STV), which corresponds to variability over days to several weeks;
and (c) long-term variability (LTV) that takes place on timescales of
months to years~\citep{Gupta2004A}.

These variabilities have many explanations that are related
to whether the source is jet-dominated or not.
In the first case, \citet{Marscher1991} suggest that the
variability in compact jets is justified because of the non-thermal emission in
blazars. Furthermore, variability studies in the radio and optical wave
bands by \citet{Camenzind1992A} conclude that shock waves  in relativistic
collimated flows are responsible for the observations. This idea was
further studied by  \citet{Mohan2015ApJ}, proposing
a general relativistic model of jet variability in AGN,
incorporating orbiting blobs in a helical motion along a magnetic surface
near the black hole. In this direction, \citet{delacruz2021} interpreted
that shock wave emissions inside the jet are caused by a periodic
injection velocity of the flow at the base of the jet.

For non-jet-dominated sources, \citet{McHardy2023}
suggest that the differences in lag between different bands indicate
that the variability is produced by the disk. Also, \citet{Edelson2002}'s
account of rapid luminosity changes that indicates emission regions confined
to the inner disk or corona. In this sense, \citet{Uttley2005} suggest that
the variability is due to processes and accretion rate variations. Also,
\citet{Stella1998ApJ} explain that the variability
 can be attributed to the relativistic dragging of inertial frames around a
rapidly rotating disk.

A famous example of a non-jet-dominated source
is represented by the quasar OJ~287, which was extensively studied
by~\citet{Lehto1996}, who found sharp flares within major outbursts of
the optical light curve.  The authors proposed a model in which a smaller
black hole crosses the accretion disk of a larger black hole during its
binary orbit. An extensive analysis of its optical light curve was used to
infer this supermassive black hole binary system~\citep{Sillanpaa1988}. 
The periodicity of this blazar was discovered by analyzing
its historical optical light curve, which contains data from more than
100~years, showing repeated bursts at intervals of about \( 11.65\,
\text{y}\).  The best-known model of this periodicity was constructed 
by~\citet{Lehto1996}, and it consists of a primary black hole---the central engine, with a mass of \(\sim\)\(17 \times 10^{9}~\text{M}_\odot\), surrounded by an accretion disk
and a secondary black hole with a mass of \(\sim\)\(10^{8}\, \text{M}_\odot \),
orbiting the primary and intersecting the accretion disk on each
orbit, causing tidally induced mass fluxes from the accretion disk to the primary
black hole.

Of particular relevance to the studies carried out in the present article
is the case for postulating the existence of a secondary supermassive
black hole orbiting a primary one.  The first proposal of this
kind was made by~\citet{Begelman1980} in order to account for periodic
or quasi-periodic oscillations.

An extensive survey to find periodic light curves in optical light
was carried out by~\citet{Graham}.  Of the  \( 247,000 \) studied light
curves, it was found that the one corresponding to PG~1302-102 shows a
periodicity of \( 1884 \pm 88\, \text{d} \).  The authors assumed that
this was due to the existence of a secondary black hole ``eclipsing''
the primary one, and they concluded that the system is separated by less than
a parsec. More recently, \citet{Tavani2018} found 2.2 yr 
QPOs in the $\gamma$$-$rays band of the blazar PG~1553+
113 and once again
proposed it as a supermassive black hole binary system.

Relevant periodicities of other AGN appear in the literature.  Quite
important to mention is the work by ~\citet{Li2016ApJ}, who report
long-term variability of \(\sim\)14~yr in the optical continuum of the
nucleus of NGC~5548. For this same object, \citet{Bon2016ApJS} found a
periodicity \(\sim\)43~yr. Also,~\citet{Li2019ApJS} studied a possible
{\ensuremath{\sim}}20~yr periodicity in long-term optical photometric and
spectral variations of the nearby radio-quiet Active Galactic Nucleus
Ark~120, and~\citet{Chen2022} published a sample of quasar candidates
with periodic variations from the Zwicky Transient Facility.

 A very interesting and quite well studied blazar is Markarian 501
(Mrk~501). It is a BL~Lac object with several periodicities reported in
the literature: (1) A periodicity of \( 23 \, \text{d} \) was reported
during a flare detected in X$-$rays and $\gamma$$-$rays and modeled
as a supermassive binary black hole~(see, e.g., \citep {Rieger2000A} and references
therein). (2) In the same frequencies, a periodicity of \(
72 \, \text{d}\) was found by~\citet{Rodig2009A}. (3) A periodicity of \(
630 \, \text{d} \) was discovered by~\citet{Wang2017Ap} in X$-$rays. (4)
Finally, \citet{Bhatta2019} found a \( 332 \, \text{d} \) periodicity
in the \textit{Fermi}
-LAT \( \gamma \)$-$rays light curve. All of these
reported periodicities are not achromatic, and they represent different
databases in time.  Although interesting, they are not relevant to the
study presented in this article, which used long-term databases in four
different frequencies.

Mrk~501 has a redshift of $z = 0.034$ ($\sim$$456 \, \text{Mly}$$\sim$$140
\,\text{Mpc}$) with R.A. $= 16^\text{h} \ 53^\text{m} \ 52.2^\text{s}$,
Dec. = $+39^\circ \, 45' \, 37''$. It was discovered by \citet{Quinn1996}
using  the Whipple Imaging Atmospheric Cherenkov Telescope
(IACT). It has been monitored since 1996 in various frequencies:
radio~\citep{Richards2011ApJ}, optical~\citep{smith0912.3621}, and
$\gamma$$-$rays~\citep{Abdo2011ApJb,Dorner:2017LT}.

In this article, we report a mean periodicity of \( 224.07 \pm
0.22 \, \text{d}\) in multi-frequency observations of Mrk~501. The
dataset in radio was obtained from the Owens Valley Radio Observatory
(OVRO) (
\url{https://sites.astro.caltech.edu/ovroblazars/}, accessed on 27 June
2020
),
the optical dataset 
from the American Association of Variable Star
Observers (AAVSO) (\url{https://www.aavso.org/}, accessed on 12 September
2021), the
X$-$rays dataset is from the Neil Gehrels Swift Observatory
(Swift) (\url{https://swift.gsfc.nasa.gov/}, accessed on 8 October 2021)
and for $\gamma$$-$rays,	the data were taken from the
\textit{Fermi} Gamma-Rays Space Telescope (FGST, also
FGRST) (\url{https://fermi.gsfc.nasa.gov/}, accessed on 13 March 2020). These datasets
and their corresponding processing (reduction) is explained
in Section~\ref{data}.	In Section~\ref{methods}, we describe
different methods to find this multifrequency periodicity, and in
Section~\ref{jacobi}, we model this periodic behavior, assuming a
supermassive binary black hole using Jacobi elliptical functions,
which offer good representations of eclipses that produce occultations
(as they occur 
for binary stars or exoplanets eclipsing their central
star) and magnifications (such as the ones that are produced by massive
relativistic objects that bend light and magnify its intensity).  Finally,
in Section~\ref{discussion}, we discuss our results.

\section{Data and Light Curves} \label{data}

The radio dataset from 22 January 2009  to 27 June 2020 was obtained
from the OVRO database.  OVRO consists of a \( 40 \, \text{m} \)
telescope with	a cryogenic receiver at a central \( 15 \, \text{GHz}
\)  frequency,	a \( 3 \, \text{GHz} \) bandwidth and two symmetric
off-axis beams. This observatory has been monitoring blazars since
2008 \citep{Richards2011ApJ}, and one of its main targets is the search
for QPOs and correlations between radio and $\gamma$$-$rays in blazars
\citep{abdo09,Ackermann2011ApJ}. The signal-to-noise level
reported by the OVRO database is such that it produces a systematic flux
uncertainty of about 5\%.

The optical AAVSO database is public, and it offers long-term datasets. The institution 
is 
an international organization of variable star observers who
participate in scientific discovery through variable-star astronomy. It
was founded in 1910, and its observations of variable stars are collected
and archived for worldwide access in collaboration with amateur and
professional astronomers. Observations with errors
of \(\geq\)1.0 magnitudes are rejected by the AAVSO community.  An error
of 1.0 magnitude represents a signal-to-noise ratio of 1, making it
statistically insignificant. The light curve of Mrk~501 was built
using the database from 24 June 1998 
to 12 September 2021.

The optical and
radio data reductions were processed via AAVSO and OVRO, respectively.
The data were obtained by consulting the public databases of both
observatories. For details of the reduction, calibration,
correlation processes, etc., of the radio database, see~\citet{Richards2011ApJ},
and for AAVSO, see~\citet{Kinne2012}.

For the X$-$ray light curve, we used the Swift database from 
2 October 2008 to 8 October 2021. This dataset contains energies in the range of
\( 0.3-10 \, \text{keV} \). Swift has an X$-$ray telescope (XRT) with
two important characteristics that makes it important for observations:
a low background and a constant point spread function across the field
of view~\citep{Moretti2005SPIE}.

The bin size used for obtaining the data was the 
default time of 5 s, and
the command line interface used was  \texttt{xselect}, which works in
conjunction with the Fermitools software version v11r5p3, 
specifically designed for
astrophysical X$-$ray analysis. It offers convenient functions to organize
input data using the observation catalog, applies various filters to the
event data (such as intensity, phase, region, etc.) and creates good
time intervals based on a chosen selection criteria. It serves as a valuable
tool to work with X$-$ray data, providing an efficient
and customizable way to manage and analyze astrophysical data.

In addition to its major functions, \texttt{xselect} 
also provides
 three commands that correspond to different time systems:
universal time (UT), modified Julian days (MJD), and the SpaceCraft
Clock. We constructed the X$-$ray light curve using MJD.

The \textit{Fermi} public database of $\gamma$$-$rays
has fluxes in the range of \( 100 \, \text{MeV}\)--\(300 \, \text{GeV}
\). For Mrk~501, it covers a time interval from 4 August 2008  up to 13 March 2020. The analysis was performed using  the public Fermi/LAT data 
corresponding to the P8R3 SOURCE  \(\gamma\)$-$ray event selection within a 15-degree
range of search. To ensure data quality, only events corresponding
to good time intervals with DATA\_QUAL  > 0 and LAT\_CONFIG == 1
were retained, and a maximal telescope zenith angle of 90 degrees
was applied. Data reduction was performed using the Fermitools package
v2.0.8. Galactic and extragalactic diffuse emission was taken into account
using the gll\_iem\_v07.fits and iso\_P8R3\_SOURCE\_V2\_v1.txt models,
respectively. The spectral shape was assumed to follow a LogParabola
model
(\url{https://fermi.gsfc.nasa.gov/ssc/data/analysis/scitools/source\_models.html}, \mbox{accessed on 13 March 2020}).

The data from the Swift and \textit{Fermi} observatories were reduced by
processing the files corresponding to both the photons and the position
and the orientation of the satellite on Flexible Image Transport
System-format files (FITS) with HEASoft version 6.26.1
 (\url{https://heasarc.gsfc.nasa.gov/docs/software/lheasoft/}, accessed on 13 March 2020) and
Fermitools version v11r5p3 
 (\url{https://fermi.gsfc.nasa.gov/ssc/data/analysis/software/}, accessed
 on 13 March 2020)
software.  In both cases, it is necessary to know the relevant
parameters of the object in question, such as the right ascension,
declination, energy interval, and starting and ending times of the
data processing~(see, e.g., \citep{gustavo}). The satellite information
was reviewed to make various time and position corrections, and with
this, the analysis of the maximum likelihood estimation was carried out,
in which the FITS files were 
used~(see, e.g., \citep{nacho}). Finally,
a table was built that contains the light curve information, i.e., a file
that contains discrete data of time, luminosity and its uncertainty
with a signal-to-noise ratio \(<\)2\(\sigma\).

\begin{figure}[H]


\includegraphics[width=6cm,
height=4.3cm]{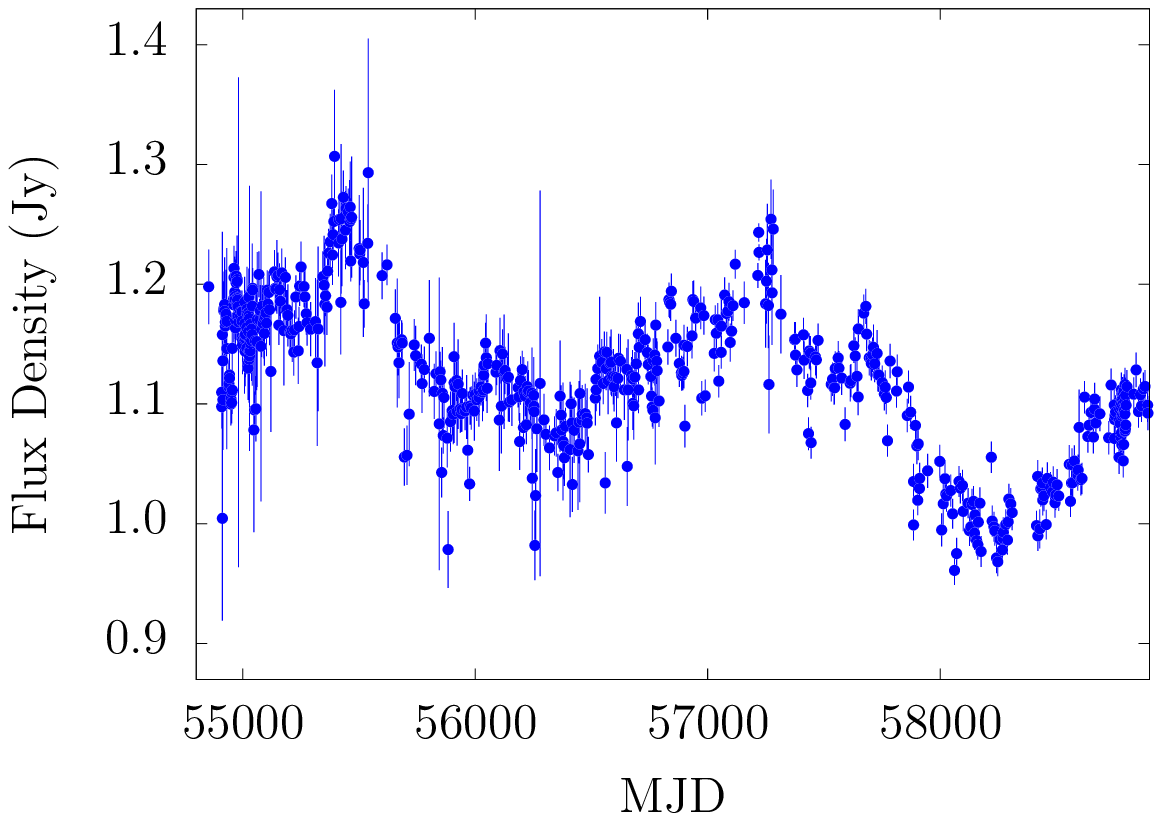}\hspace{1cm}
\includegraphics[width=6cm, height=4.3cm]{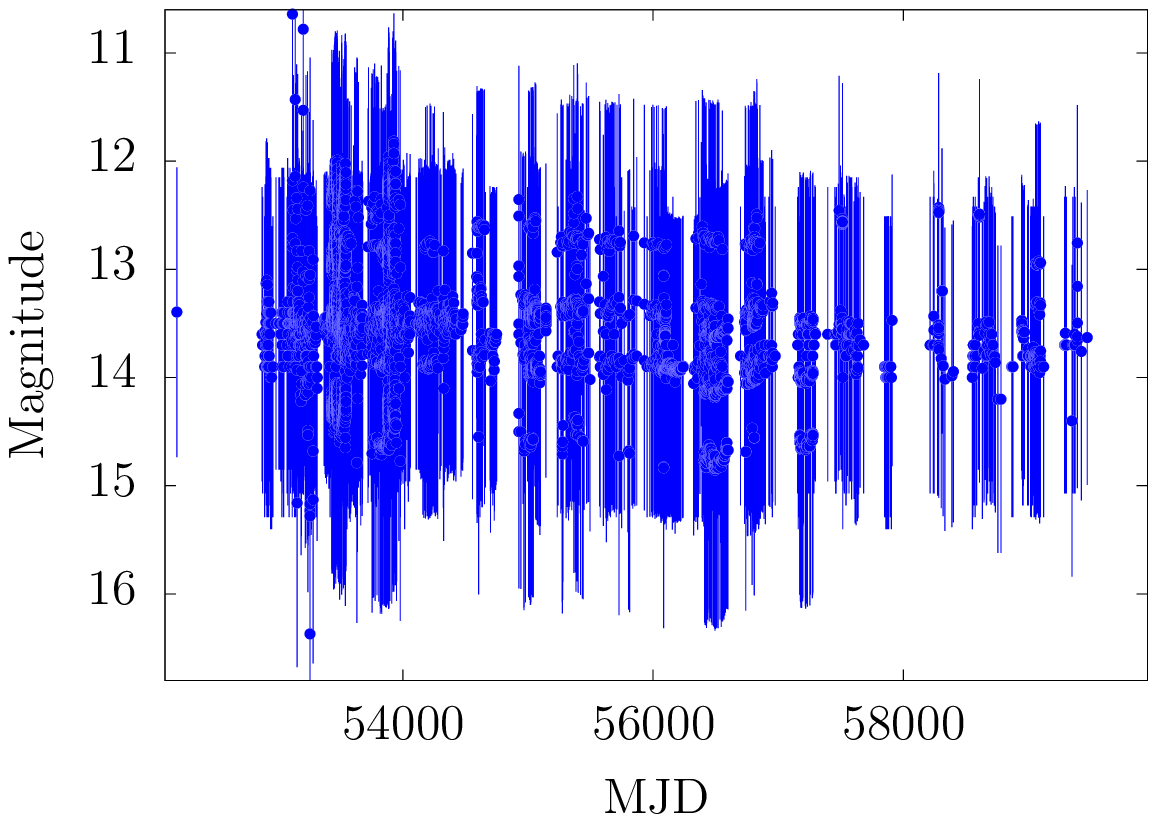}\\
\vspace{.5cm} \includegraphics[width=6cm,
height=4.3cm]{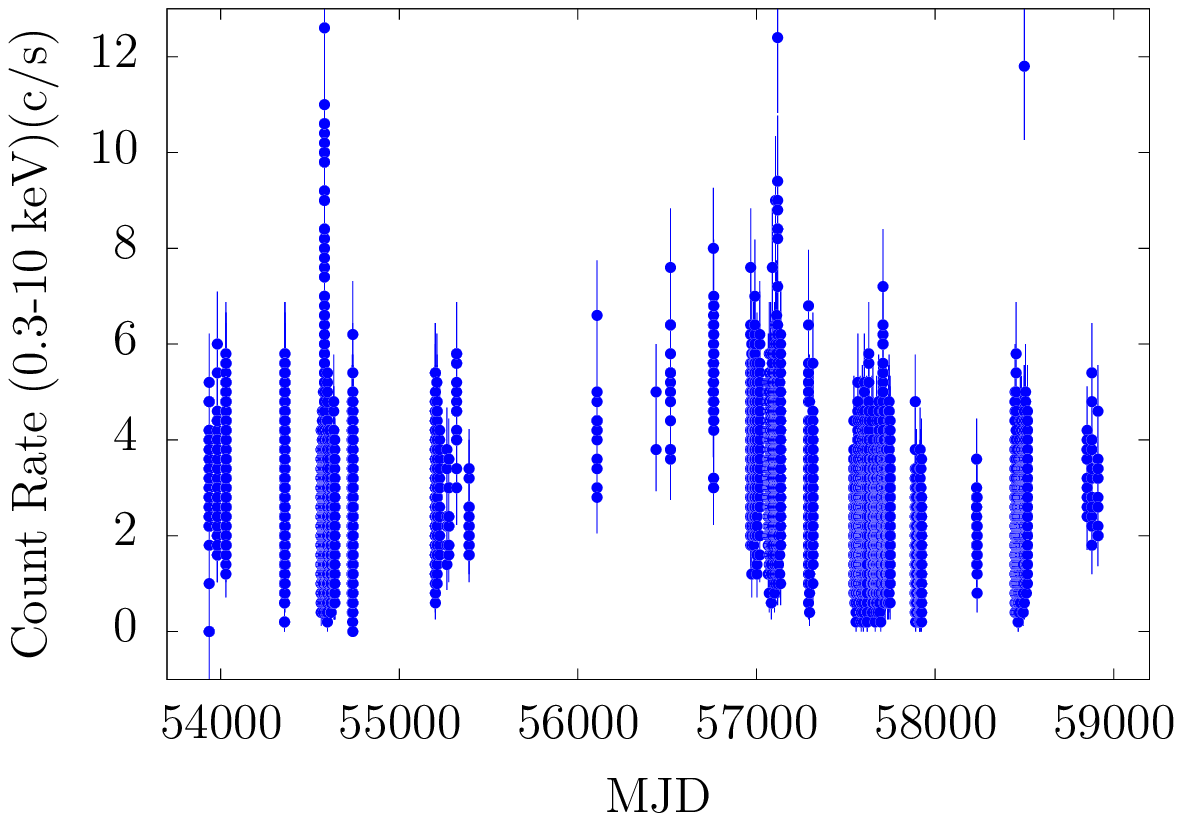}\hspace{1cm}
\includegraphics[width=6cm, height=4.3cm]{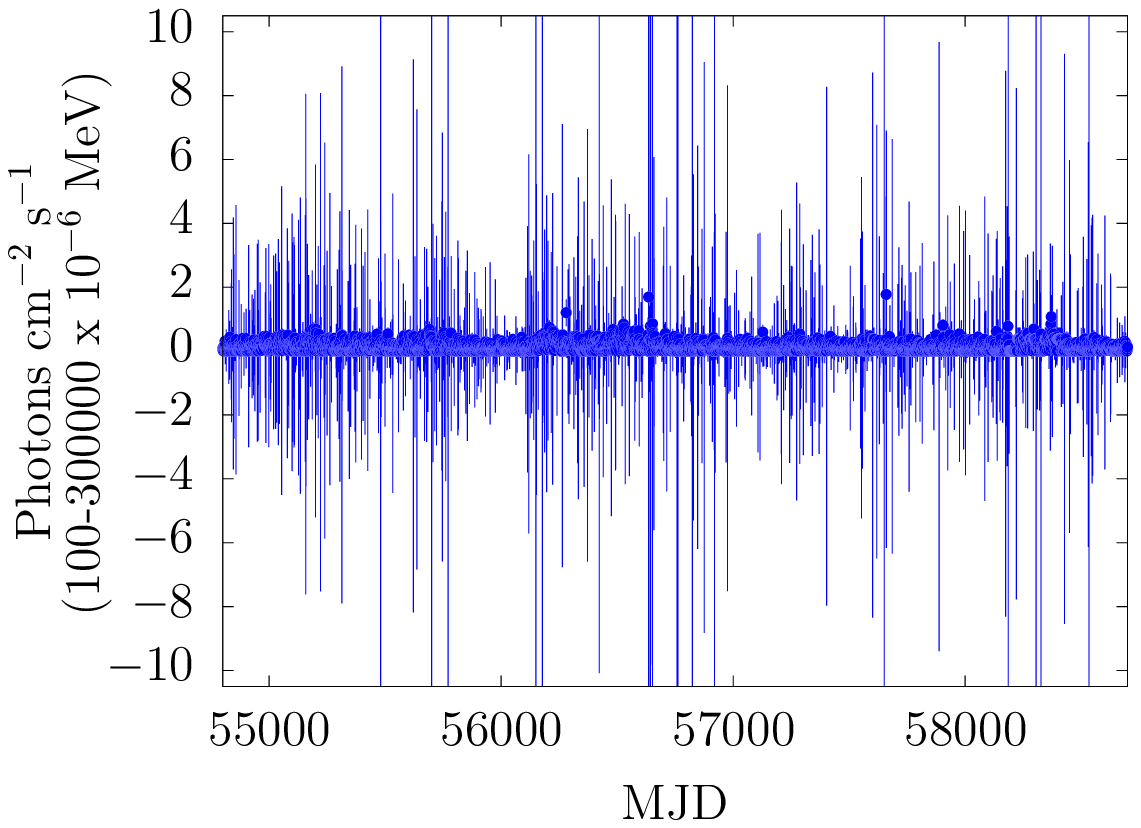}\\
\caption{Multifrequency 
 light curves of the blazar Mrk~501.
From left to right and top to bottom, the panels represent radio, optical,
X$-$ and $\gamma$$-$rays at a \( 3\sigma \) confidence level for 
light curves, as
described in \mbox{the text.}}

\label{fig-lc}

\end{figure}

The multifrequency light curves in Figure~\ref{fig-lc} 
are the result of processing the obtained data described above at 
a \( 3\sigma \) confidence level; thus,
the accepted total fraction of bins is 99.7\%. 
Table~\ref{tab:datos-mrk501-b} shows the processed data in a synthesized
way for all the electromagnetic frequencies studied for Mrk~501.


\vspace{-6pt}



\begin{table}[H]

\caption{The 
table shows each electromagnetic band studied for Mrk~501:
the number of observations, the duration in years,
months and days (y, m, d), and the total days for which at least a
measurement was obtained (d).}




\setlength{\cellWidtha}{\columnwidth/4-2\tabcolsep-0.0in}
\setlength{\cellWidthb}{\columnwidth/4-2\tabcolsep-0.0in}
\setlength{\cellWidthc}{\columnwidth/4-2\tabcolsep-0.0in}
\setlength{\cellWidthd}{\columnwidth/4-2\tabcolsep-0.0in}

\scalebox{1}[1]{\begin{tabularx}{\columnwidth}{>{\PreserveBackslash\centering}m{\cellWidtha}>{\PreserveBackslash\centering}m{\cellWidthb}>{\PreserveBackslash\centering}m{\cellWidthc}>{\PreserveBackslash\centering}m{\cellWidthd}}
\toprule

\textbf{Band}  &  \textbf{Observations} & \textbf{Duration}
& \textbf{Total}\\

\phantom{AA} & \phantom{AA} & \ \textbf{(y, m, d)} &\textbf{(d)}\\

\cmidrule{1-4} 

Radio  &  \( 615 \) &  \(11, 5, 5 \) & \( 4174	\) \\

Optical  & 11,849  & \(23, 2, 17 \) & \( 8441  \)	\\

X$-$rays	&  28,000  &  \( 12, 7, 9 \) & \( 4607\)  \\

$\gamma$$-$rays &  {4199} 
 & \( 11, 7, 9 \) & \( 4239  \)  \\

\bottomrule

\end{tabularx}}

\label{tab:datos-mrk501-b}


\end{table}


\section{Methods} \label{methods}

\subsection{Periodograms and Window Functions} \label{periodograms}

With the data processed, the search for periodicities was carried
out. For this, the R-package RobPer version 1.2.3 was useful to analyze the
associated periodograms on scales of a time of several years. This
package was built for the analysis of time series in astrophysics, in
which there is the possibility of having data that are not necessarily
equidistant in time~\citep{anita02}.  It is based on robustly fitting
periodic functions of the time-series (light curves) in question and
calculates periodograms with irregular (non-equidistant) observations
on the time scale. The RobPer routines report significant peaks on the
periodograms based on different statistical techniques~(see \citep {anita02} and
references therein), with regression options
such as least squares, least absolute deviations, least trimmed squares,
and M-, S- and \(\tau\)-regression 
while accounting for measurement
accuracies with weights. RobPer covers previous approaches and introduces
model-regression combinations.	Instead of relying on fixed critical
values, it employs an outlier search on the periodogram to identify
valid periods, considering that the traditional assumptions might not
hold. RobPer, being a robust technique, is less sensitive to drastic
changes in small portions of the data, making it more robust \mbox{to 
outliers.}

To support the RobPer periodogram analysis, a Lomb--Scargle (L-S)
periodogram was also used, applying the lomb statistical routine in R
and the LombScargle class of the astropy.timeseries routine in Python.
This algorithm was proposed by~\citet{Lomb1976Ap} and~\citet{Scargle1982},
and to date, it is the most-used method for detecting periodicities in
unevenly sampled light curves.	The L-S technique is based on Fourier
methods, phase-folding methods, least squares methods and Bayesian
approaches~\citep{VanderPlas2018}. Despite the success of the L-S
routine, it is well known that it fails in some cases where eclipses
are to be detected as periodicities~(see, e.g., \citep{baluev15}).
This means that one has to proceed with care when studying the L-S
periodogram. In any case, not a single statistical method presented in
this article can, in general, serve as a unique tool for the detection
of periods on a time series. This is the reason why we studied
the problem using different statistical techniques.

  To avoid false positives in the detected periods with the RobPer
and L-S routines, we built a  windowing program following the work
of~\citet{dawson10} and applied it to all the electromagnetic bands studied.

This method effectively distinguishes between aliases and
true frequencies. Aliases occur at  \(|f \pm f_{s}|\), where f represents
the frequency, and \(f_{s}\) is a feature in the window function. By
comparing the phase and amplitude of predicted aliases, derived from a
sinusoidal waveform with the candidate frequency and the actual data,
it is possible to determine whether the predicted aliases align with the
data. When all aliases match in terms of amplitude, phase and pattern,
we can confidently conclude that the true period has been found.

Thus, the spectral window function of an evenly sampled time series is

\begin{equation} 
  W(\nu) = \frac{1}{N}\sum_{r=1}^{N}e^{-2\pi i\nu t_{r}}
\end{equation}

\noindent with \( N \)  
as the number of data points,  \(m\)
is an integer and  \(t_{r}\) is the time of the  \(r\)th data point. The
peak in the spectral window function occurs at  \(mf_{s}\).

\begin{figure}[H]
  \includegraphics[width=6cm,
  height=4.3cm]{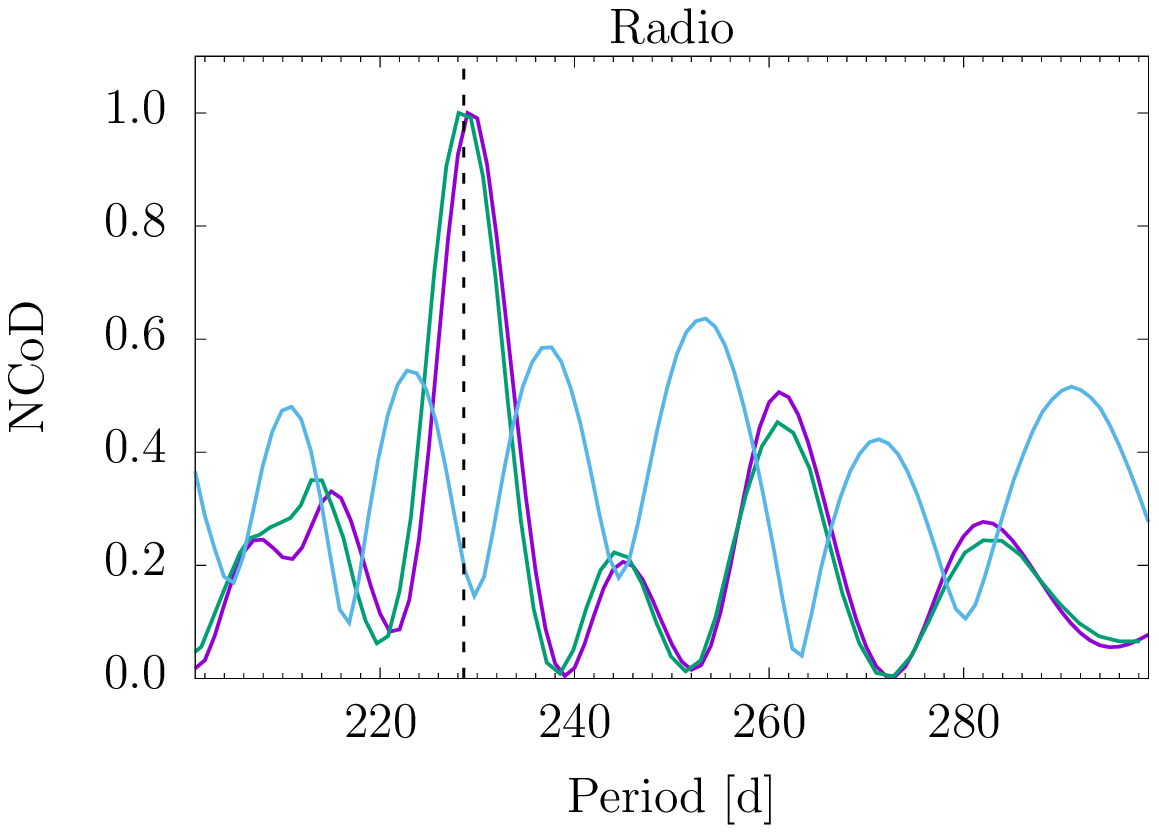}\hspace{1cm}
  \includegraphics[width=6cm,
  height=4.3cm]{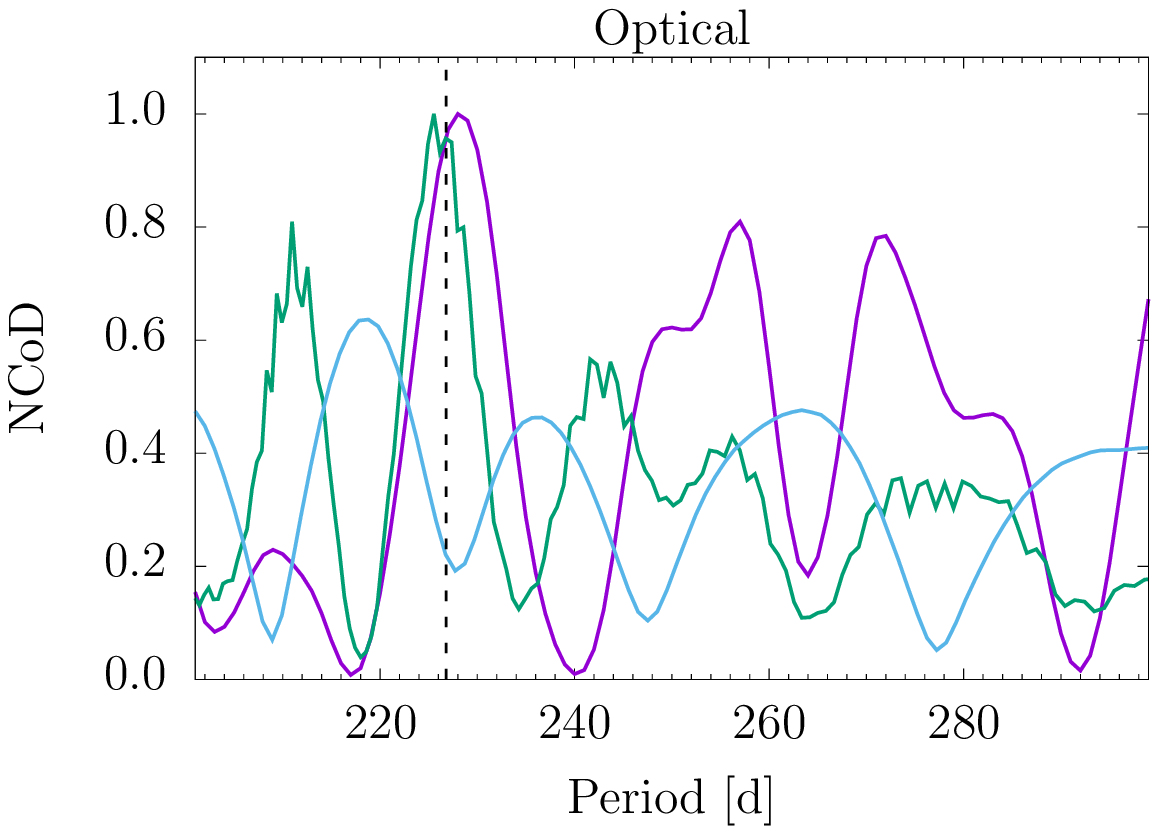}
  \\ \vspace{.5cm} \includegraphics[width=6cm,
  height=4.3cm]{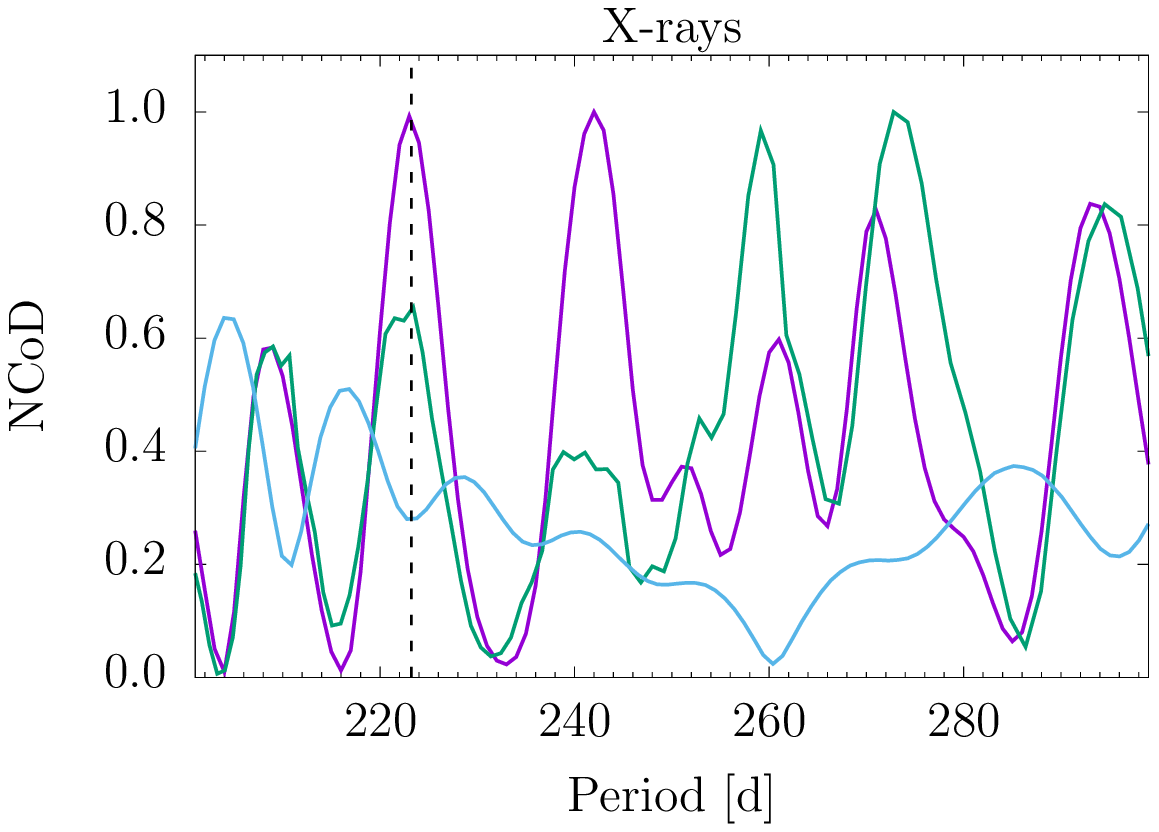}\hspace{1cm}
  \includegraphics[width=6cm, height=4.3cm]{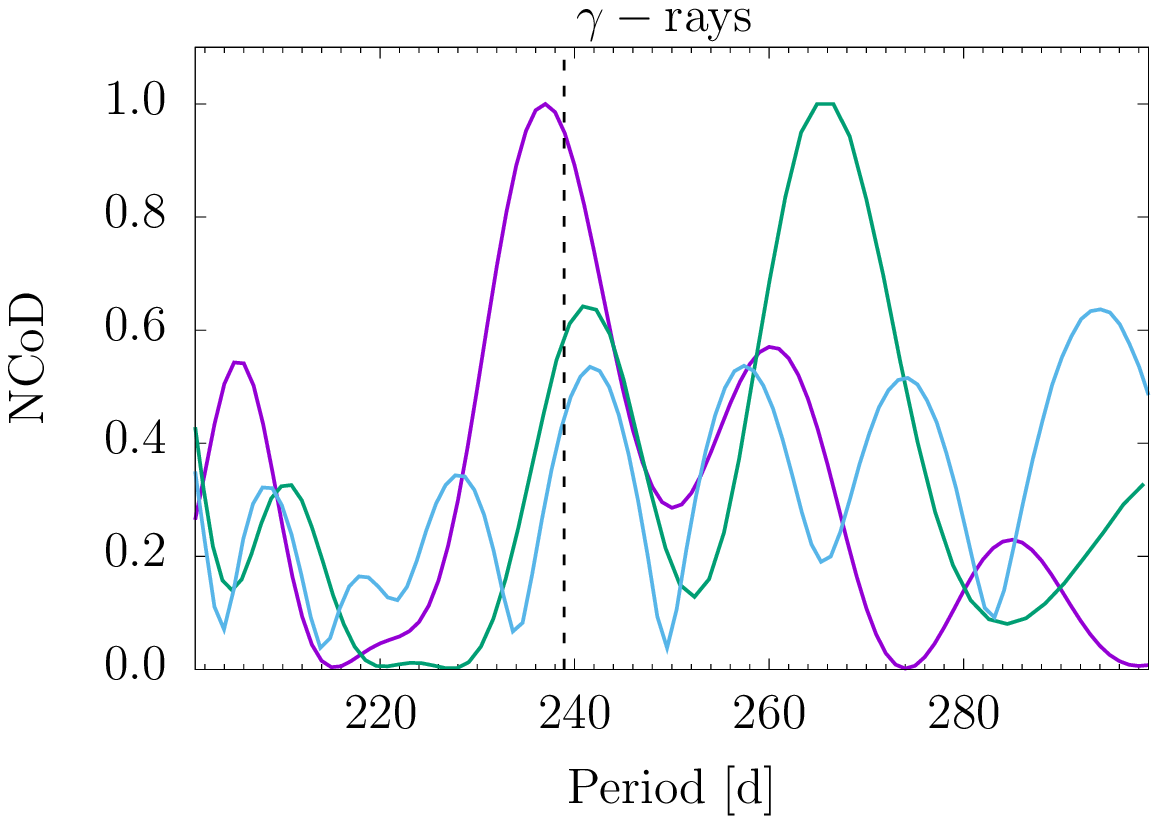}
  

\caption{RobPer (magenta) and L-S (teal) periodograms (represented by
their normalized coefficient of determination -NCoD), together with their
corresponding window function (blue) for radio-, optical-, X$-$ and
 \(\gamma \)$-$ray observations of Mrk~501 are shown from left to right,
top to bottom panels. The black dotted vertical line shows the mean
periodicity between the RobPer and L-S peaks that are common
in all frequencies (radio: \(228.03 \, \text{d}\);
optical: \(226.77 \, \text{d}\); X$-$rays: \(223.20 \, \text{d}\);
and \(\gamma \)$-$rays: \(238.90 \, \text{d}\)). Table~\ref{tab:per}
shows the periods obtained for the periodograms presented in the figure.
The fact that these mean periodicities do not coincide with peaks in
the window function reinforces their true \mbox{periodic character.}  }
\label{fig-per}
\end{figure}

Under this method, false positives are ruled out via a direct comparison of
a $W(\nu)$ vs. $\nu$ diagram and the periodograms obtained with RobPer
and L-S. In this way, if there is a false positive periodicity in the
data collection (such as, e.g., an annual period of time due to vacation
days in which no observations were taken), it will be presented as a peak
in the window function (WF). The peaks that result as true positives
in a periodogram are, thus, compared with their respective WF troughs in
all the electromagnetic bands analyzed. Figure~\ref{fig-per} shows the
periodograms obtained with RobPer and L-S with the corresponding time, WF.
The true mean peak in each periodogram is shown with a vertical dotted
line. These mean periodicities are reinforced by the fact that the true
mean peaks do not coincide with any peak in the WF.  Figure~\ref{fig-per}
only shows a zoom into the relevant part of the periodogram, where a
peak on it approximately coincides in all studied frequencies.


\vspace{-6pt}


\begin{table}[H]
\caption{The table shows the time in days
(d) for the peaks and their mean values for the RobPer and Lomb--Scargle
algorithms in each electromagnetic band studied for Mrk~501.}
  
  \setlength{\cellWidtha}{\columnwidth/4-2\tabcolsep-0.0in}
\setlength{\cellWidthb}{\columnwidth/4-2\tabcolsep-0.0in}
\setlength{\cellWidthc}{\columnwidth/4-2\tabcolsep-0.0in}
\setlength{\cellWidthd}{\columnwidth/4-2\tabcolsep-0.0in}

\scalebox{1}[1]{\begin{tabularx}{\columnwidth}{>{\PreserveBackslash\centering}m{\cellWidtha}>{\PreserveBackslash\centering}m{\cellWidthb}>{\PreserveBackslash\centering}m{\cellWidthc}>{\PreserveBackslash\centering}m{\cellWidthd}}
\toprule

\textbf{Band} &	\textbf{RobPer} &  \textbf{L-S} & \textbf{Mean}\\ 
\phantom{kk} & \textbf{(d)} & \textbf{(d)} & \textbf{(d)}\\ 
\cmidrule{1-4} 
Radio & \( 228 \) & \( 228.06 \)  & \( 228.03\) \\ Optical  & \( 228 \)
& \( 225.54 \)	& \( 226.77\) \\ X$-$rays  & \( 223 \) & \( 223.40 \) &
\( 223.20 \) \\ $\gamma$$-$rays  & \( 237 \) & \( 240.80 \) & \( 238.90
\) \\ 
\bottomrule
\end{tabularx} }

\label{tab:per} 
\end{table}


\subsection{VARTOOLS} 
\label{vartools}

VARTOOLS (The VARTOOLS free GNU
General Public License software is available a:
\url{https://www.astro.princeton.edu/\~jhartman/vartools.html}, accessed on 6 August 2023) is an
open-source command-line utility for analyzing light curves developed
by~\citet{Hartman2016A}. It is written in the C programming language, and it provides a set of tools for processing, manipulating and studying
light curves. The routines implemented in VARTOOLS for studying Mrk~501
were the L-S periodogram, the box-least squares (bls), the analysis of
variance (aov) and the discrete Fourier transform (DFT).

AoV is a VARTOOLS routine for the detection of sharp (or statistically
significant)
  periodic signals based on the code developed
by~\citet{Devor2005ApJ}. This method consists of folding
and binning data with a trial period. It has also been used
by~\citet{Schwarzenberg-Czerny1989MNRAS} for identifying and
studying eclipsing binaries within large data sets of light curves.
This routine customizes the process with a minimum period
(\texttt{minp} option) and a maximum period (\texttt{maxp} option) to
define the search range. The initial search uses a frequency resolution
of the subsample, and the refined peak periods utilize a finetuned resolution. 
The program outputs the highest peaks in the periodogram.

The AoV-h  VARTOOLS routine consists of the analysis variance using a multi-harmonic model. This method uses
periodic orthogonal polynomials to fit the observations and the
analysis of the statistical variance to evaluate the quality of the
fit~\citep{Schwarzenberg-Czerny1996ApJ}. The parameters
and behavior are as for the AOV routine.

\begin{figure}[H]
\includegraphics[width=6cm,
  height=4.3cm]{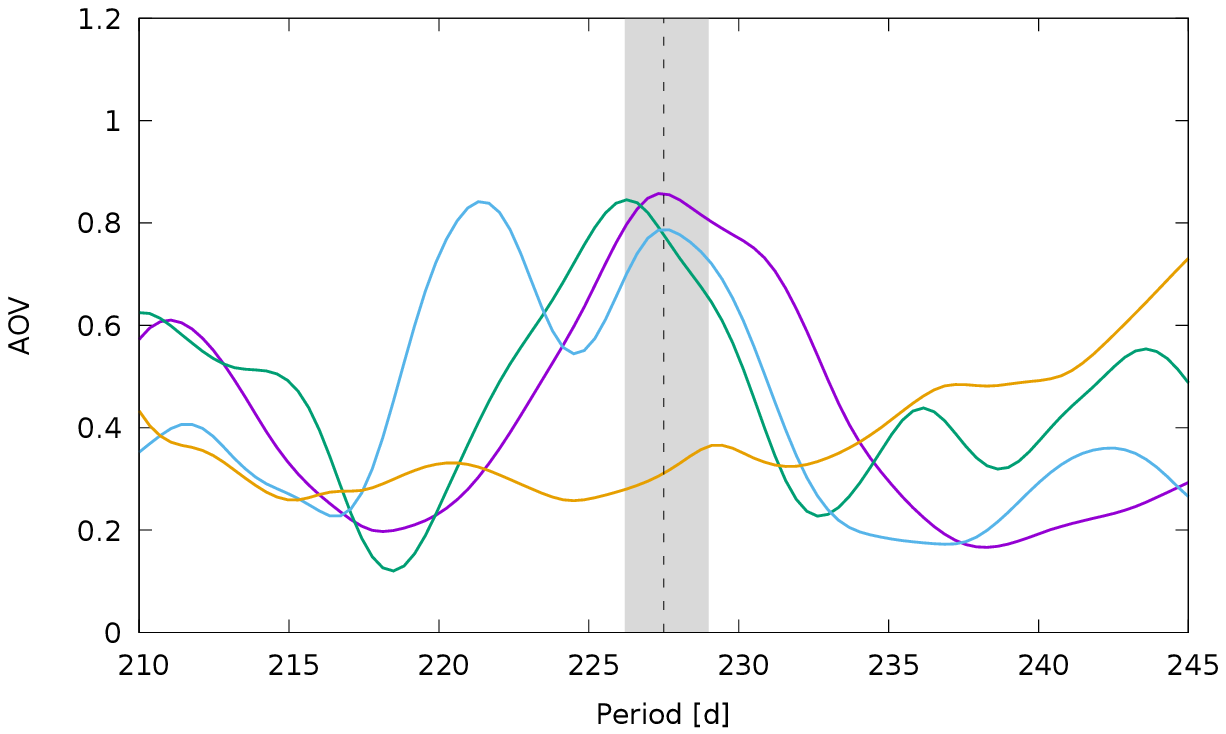}\hspace{1cm}
  \includegraphics[width=6cm, height=4.3cm]{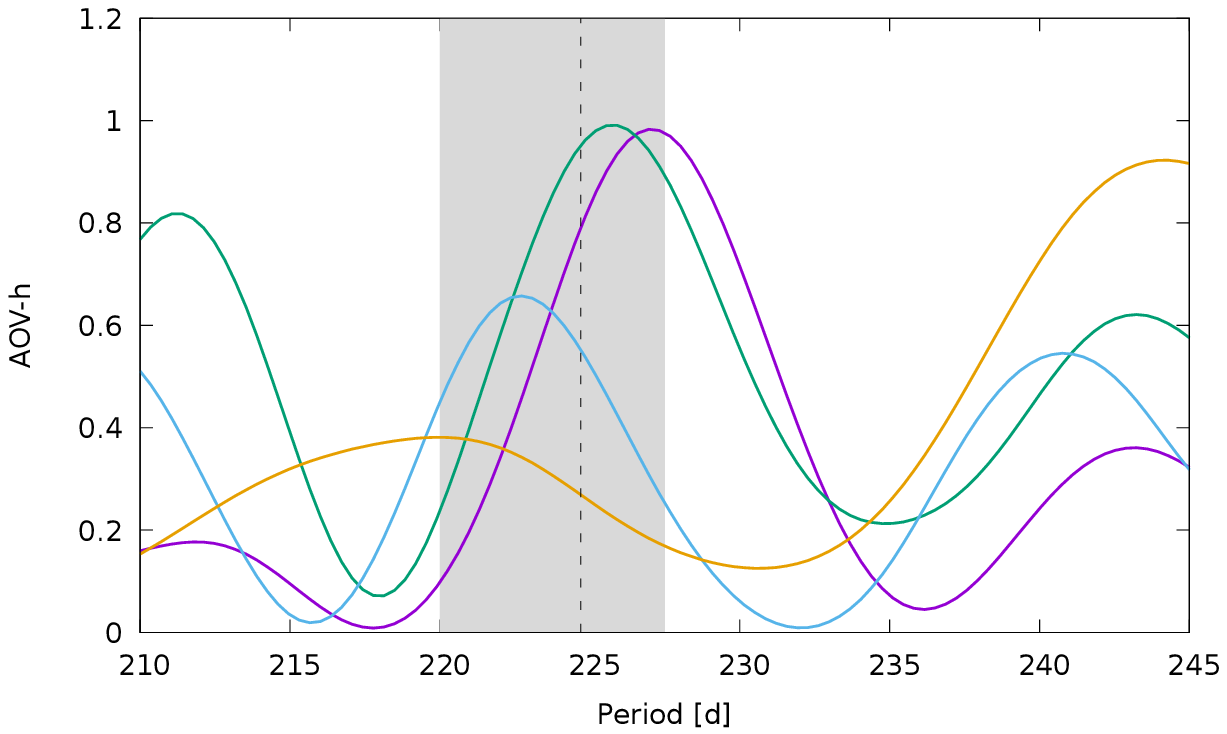} \\

\caption{The 
 figure shows the analysis of variance (AoV) for Mrk~501
using VARTOOLS.  The left panel is the AoV for all frequencies: radio is in
magenta with a periodicity of  \( 227.2 \, \text{d} \), optical is in blue
with a periodicity of \( 226.73 \, \text{d} \), X$-$rays, in teal, have
a periodicity of \( 227.1 \, \text{d} \) and $\gamma$$-$rays, in yellow
with a periodicity of \(  229.2 \, \text{d} \). The mean value of \(
227.55 \, \text{d} \) is represented with a dashed vertical line. The
gray vertical band zone represents the statistical range of these peaks.
The right panel uses the same coloring scheme as the left one but for
the harmonic analysis of variance (AoV-h) of the {VARTOOLS software
version 1.40}
.
The periodicity of radio, optical, X$-$ and \( \gamma \)$-$rays are
given via the following: \( 228.06 \, \text{d} \),  \( 227.4 \, \text{d} \), \( 223.4
\, \text{d} \) and \(  219.7 \, \text{d} \), respectively, yielding an
average value of \( 224.64 \, \text{d} \), shown with the vertical dashed
line. The vertical values on both panels were normalized to the maximum.}

\label{fig-aov} 
\end{figure}

The plots of the AoV and AoV-h of Mrk~501 are shown in
Figure~\ref{fig-aov}, which result in an average periodicity of
\(227.55 \,  \text{d} \) for the AoV and \(224.64 \, \text{d} \)
for the AoV-h. The gray vertical zone represents the statistical range
of these peaks.


\begin{figure} [H]

  \includegraphics[width=6cm,
  height=4.3cm]{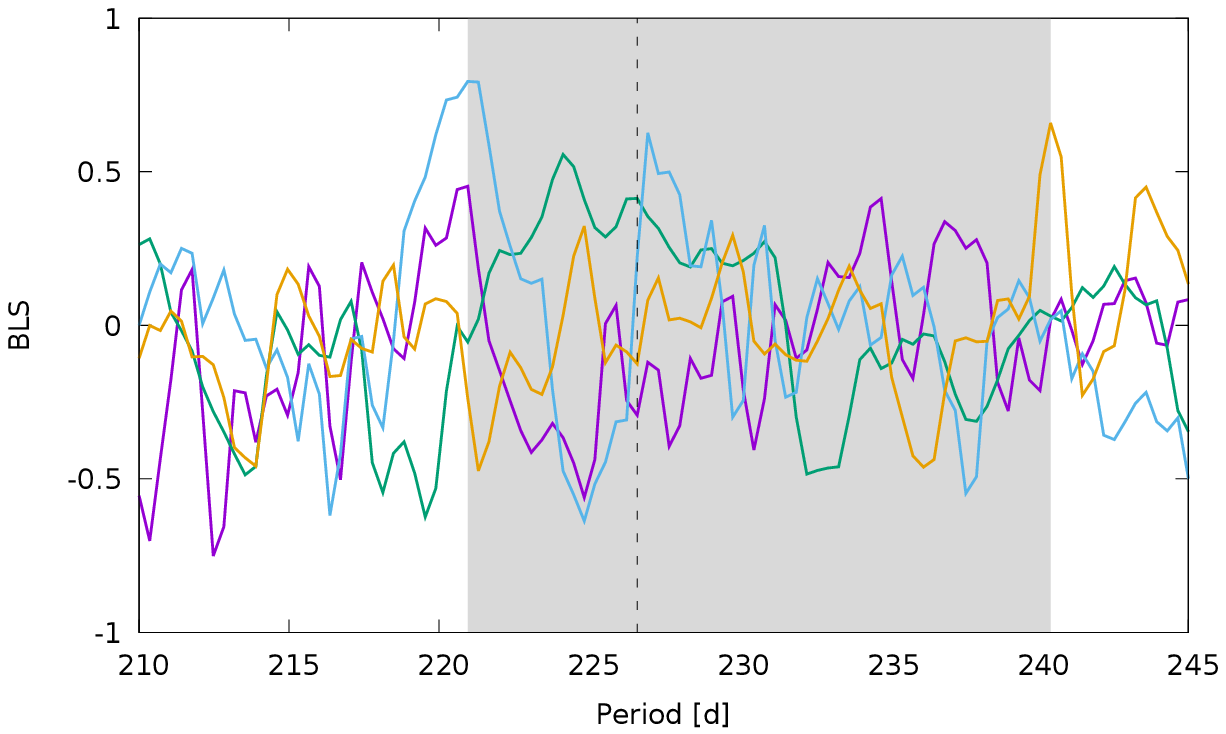}\hspace{1cm}
  \includegraphics[width=6cm, height=4.3cm]{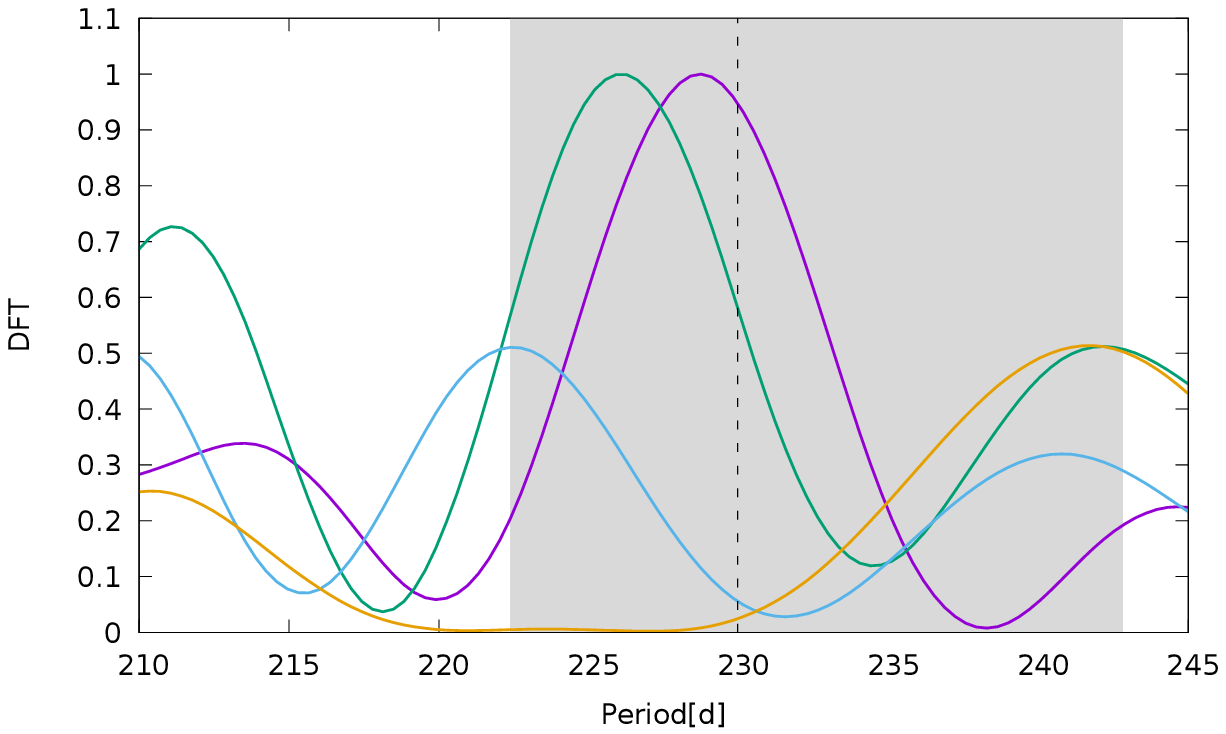}
  \\
\caption{The 
 left panel shows the B-L Square algorithm
of VARTOOLS used in all wavebands, applying the same coloring scheme
of Figure~\ref{fig-aov}, with the following periodicities in radio,
optical, X$-$ and \( \gamma \)$-$rays: \( 220.96 \, \text{d} \), \(
224.141\, \text{d} \), \( 220.96
	\, \text{d} \), \( 240.404 \, \text{d} \), with a mean value of
	\(226.616 \,
\text{d} \) represented using a dashed horizontal line. The right panel, with
the same coloring scheme, uses the DFT VARTOOLS algorithm with resulting
periodicities of \( 228.737 \, \text{d} \), \( 225.909~\text{d} \),
\(222.374 \, \text{d} \) and \(242.818 \, \text{d} \) and an average
value of \(229.959 \, \text{d} \). The vertical values on both panels
were normalized to the maximum. The gray vertical band zone represents
the statistical range of these peaks. } \label{fig-bsl-dft} 
\end{figure}

The VARTOOLS BLS routine is commonly used in studies of stellar
photometric time series in the search for periodic transits of
exoplanets~\citep{Kovacs2002A}. This algorithm uses
the minimum and maximum values (the fraction of orbit in transit), and the
search for periodicities operates within specified minimum and maximum
periods (\texttt{minper} and \texttt{maxper} commands, respectively) and uses a
specified number of trial frequencies. Additionally, the command allows
the number of peaks to be reported. 
Using this routine for  Mrk~501, we found
a mean periodicity in all frequencies of \(229.484 \, \text{d} \)---see
Figure~\ref{fig-bsl-dft}. 

The VARTOOLS Discrete Fourier Transform (DFT)
algorithm calculates the power spectrum of the time
series~\citep{Roberts1987A}. The routine to compute the DFT was developed
by~\citet{Kurtz1985MNRAS}. The computation of the light
curves is sampled using points per frequency and set to the maximum
frequency (with \texttt{maxfreq} option), with a default value based
on the minimum time separation. The DFT also allows researchers to find 
the highest
peaks in the clean power spectrum. The average periodicity value obtained
with this tool is \( 229.95 \, \text{d} \) for Mrk~501 in all studied
frequencies---see Figure~\ref{fig-bsl-dft}. Table~\ref{tab:vartools}
summarizes the the periodicity results of VARTOOLS for Mrk~501.

\begin{table}[H]
\caption{The table shows the mean periodicity value of the AoV, AoV-h,
BLS and DFT obtained with VARTOOLS for Mrk~501.} \label{tab:vartools}


\setlength{\cellWidtha}{\columnwidth/5-2\tabcolsep-0.0in}
\setlength{\cellWidthb}{\columnwidth/5-2\tabcolsep-0.0in}
\setlength{\cellWidthc}{\columnwidth/5-2\tabcolsep-0.0in}
\setlength{\cellWidthd}{\columnwidth/5-2\tabcolsep-0.0in}
\setlength{\cellWidthe}{\columnwidth/5-2\tabcolsep-0.0in}
\scalebox{1}[1]{\begin{tabularx}{\columnwidth}{>{\PreserveBackslash\centering}m{\cellWidtha}>{\PreserveBackslash\centering}m{\cellWidthb}>{\PreserveBackslash\centering}m{\cellWidthc}>{\PreserveBackslash\centering}m{\cellWidthd}>{\PreserveBackslash\centering}m{\cellWidthe}}
\toprule

\textbf{Band} &  \textbf{AoV} &  \textbf{AoV-h }&  \textbf{BLS}
& \textbf{DFT}\\ 
\ \ \	& \textbf{(\text{d})} & \textbf{(\text{d})}  &  \textbf{(\text{d})} & \textbf{(\text{d})}\\
\cmidrule{1-5} 

Radio & \( 227.20 \) & \( 228.06 \)  & \( 220.960 \) & \( 228.737\)
\\ Optical  & \( 226.73\)  & \( 227.40\)  & \( 224.141\)  & \( 225.909\)
\\ X$-$rays  & \( 227.10\)  & \( 223.40\)  & \( 220.960\)  & \( 222.374\)
\\ \(\gamma\)$-$rays  & \( 229.20\)  & \(  219.70\)  & \(240.404\) &
\( 242.818\) \\  \hline Averages & \(  227.55\)  & \(  224.64\) & \(
226.616\)  & \(  229.959\)  \\ 
\bottomrule

\end{tabularx}}

 \end{table}


\subsection{Power Spectrum Density, Detrended Fluctuation Analysis,
and the Colors of Noise} \label{power-spectrum}

The study of time series or light curves, $f(t)$, in the study of this
article, with sample length $n$, can be analyzed using the Fourier
transform $P(\nu)$ given via (see, e.g., \citep{Aschwanden2011}):

\begin{equation} P(\nu) = \frac{1}{n}\sum_{t=0}^{n-1}f(t) e^{(-\frac{2\pi
  i\nu t}{n} )}.
\end{equation}

\noindent With 
 this transformation, it is possible to obtain specific
periodic pulses in the light curve, even with excessive noise~(see, e.g., \citep{Press1978}). If there are multiple periodic fluctuations present
in the light curve, then the power spectral density (PSD) will reveal
each one with a peak in the power spectrum at the particular periodicity
or frequency.

The PSD function approaches a power law at a particular frequency, $\nu$,
given via the following:

\begin{equation} P(\nu) = \nu^{-\alpha}, \label{psd-equation}
\end{equation}

\noindent for a fixed exponent, $\alpha$. The spectral noise of the
signal can be represented with its color of noise~\citep{Press1978},
depending on the value of \( \alpha \).  For astrophysical phenomena,
it is 
commonly denoted as white, pink and Brownian according
to the values of Table~\ref{tab:colornoise}~(see, e.g., \citep {Aschwanden2010} and
references~therein).

\begin{table}[H]
\caption{The table shows the intervals of the power spectrum density
(PSD) exponent \( \alpha \) of Equation~\eqref{psd-equation} and its
associated color of noise.} \label{tab:colornoise} 



\setlength{\cellWidtha}{\columnwidth/2-2\tabcolsep-0.0in}
\setlength{\cellWidthb}{\columnwidth/2-2\tabcolsep-0.0in}

\scalebox{1}[1]{\begin{tabularx}{\columnwidth}{>{\PreserveBackslash\centering}m{\cellWidtha}>{\PreserveBackslash\centering}m{\cellWidthb}}

\toprule

\textbf{PSD Exponent} &  \textbf{Colors of Noise} \\ 
\cmidrule{1-2} 

$ 0.0 \lesssim  \alpha   \lesssim  0.5$ & white \\
 $0.5 \lesssim	\alpha \lesssim   1.5$ & pink \\
  $1.5 \lesssim  \alpha \lesssim   2.5$ & Brownian \\

\bottomrule

\end{tabularx}}


\end{table}

A given color of noise has statistical and correlation characteristics.
When white noise predominates in the signal, it means that there is
no temporal correlation in a specific time series, and the case of
Brownian noise is obtained when a temporal correlation in 
the signal
is significant~\citep{Schroeder1991}. Pink noise corresponds to the
statistical case of phase change, in which there is a transition from
a random process to a predictive one~\citep{May1976}. The colors of
noise analysis has been performed in the literature for time series
of solar bursts, magnetospheric physics, binary black holes and other
astrophysical phenomena~\citep{Aschwanden2010}.

To support the PSD results, a detrended fluctuation analysis (DFA)
method~\citep{PengPhysRevE.49.1685} with the time series can be computed
to determine the statistical self-affinity in each frequency.
DFA is also useful in the analysis of time series that appear to show
long-memory processes. It is useful in the study of chaos theory,
stochastic processes and time series analysis. The fluctuation, $F(n)$,
is calculated for different window sizes, $n$, as follows:

\begin{equation} F(n) = \sqrt{\frac{1}{N} \sum_{t=1}^{N} (X_{t} -
  Y_{t})^2},
\end{equation}

\noindent where $N$ is time series length, $X_{t}$ is the cumulative
sum or profile of the time series and $Y_{t}$ is the resulting piecewise
sequence of straight-line fits~(see, e.g., \citep {bryce2012revisiting}).
The resulting $\log  F(n) $~vs.~$\log  n $ measures the statistical
self-affinity of the time series, expressed through a fixed \( \beta \)
exponent from the following relation:

\begin{equation} F(n)\propto n^{\beta }.  \end{equation}

\begin{figure}[H]


\includegraphics[width=6cm,height=4.3cm]{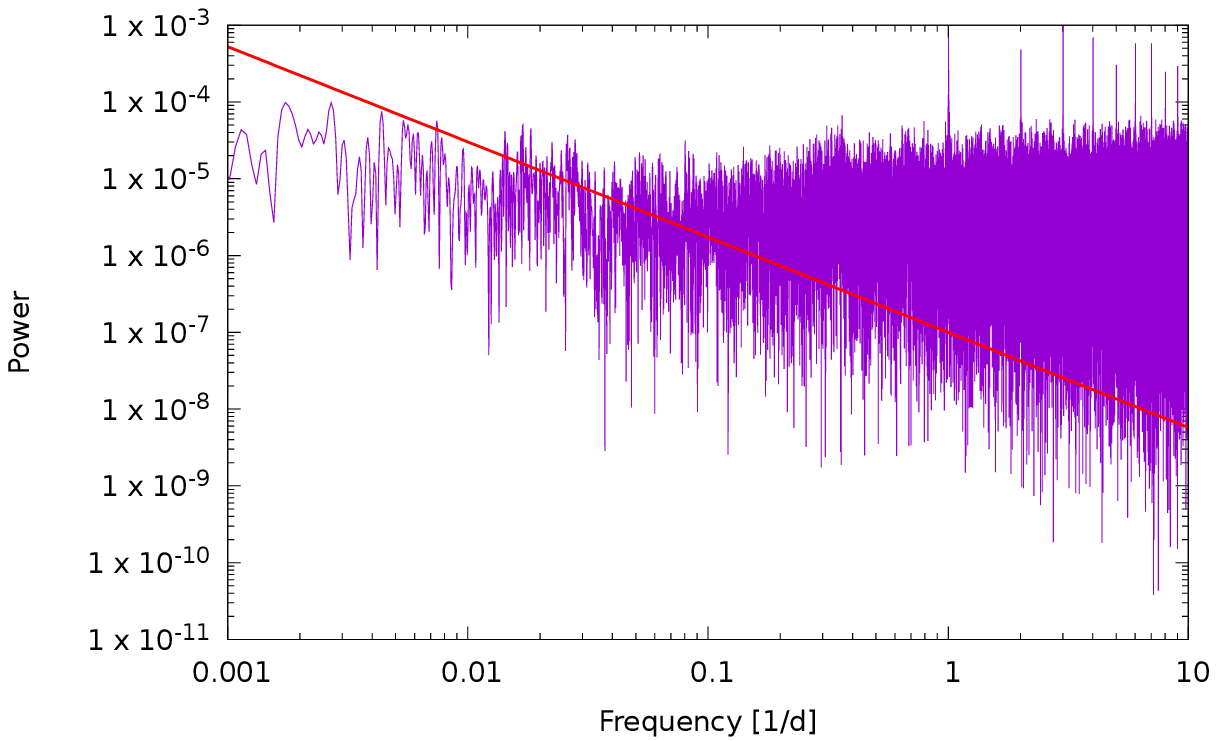}
\vspace{1cm}
\includegraphics[width=6cm,height=4.3cm]{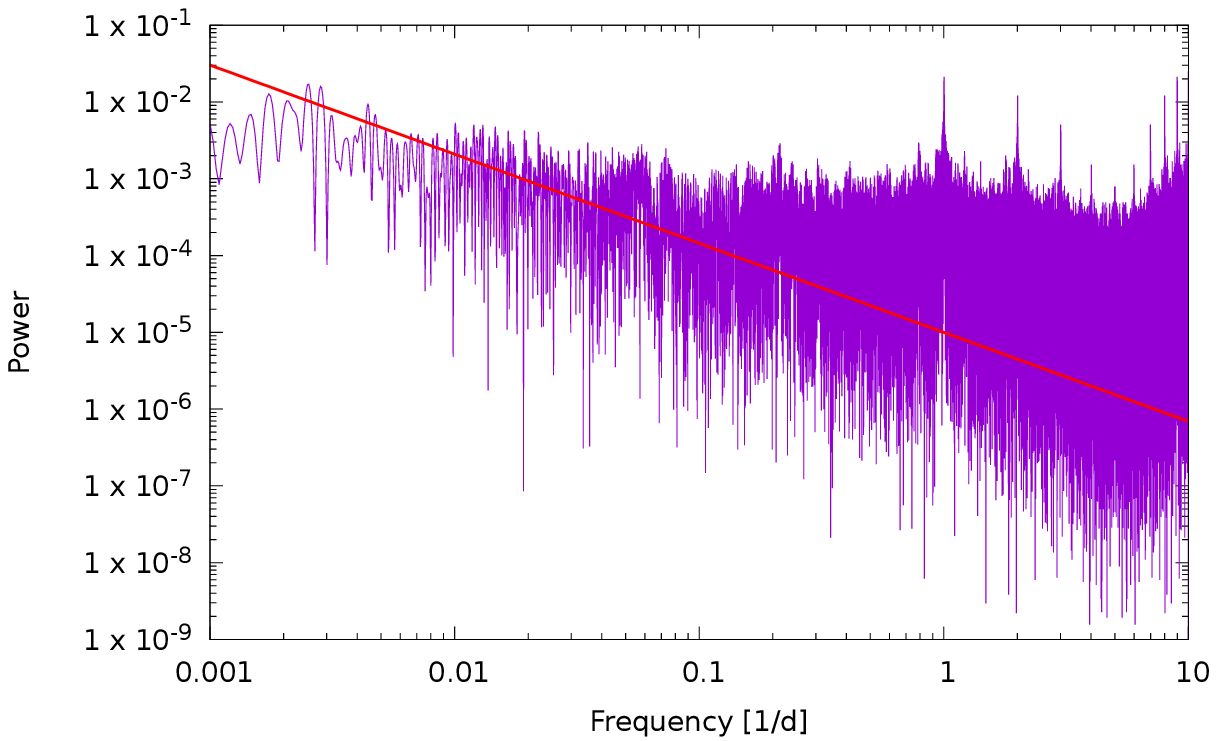}\\
\includegraphics[width=6cm,height=4.3cm]{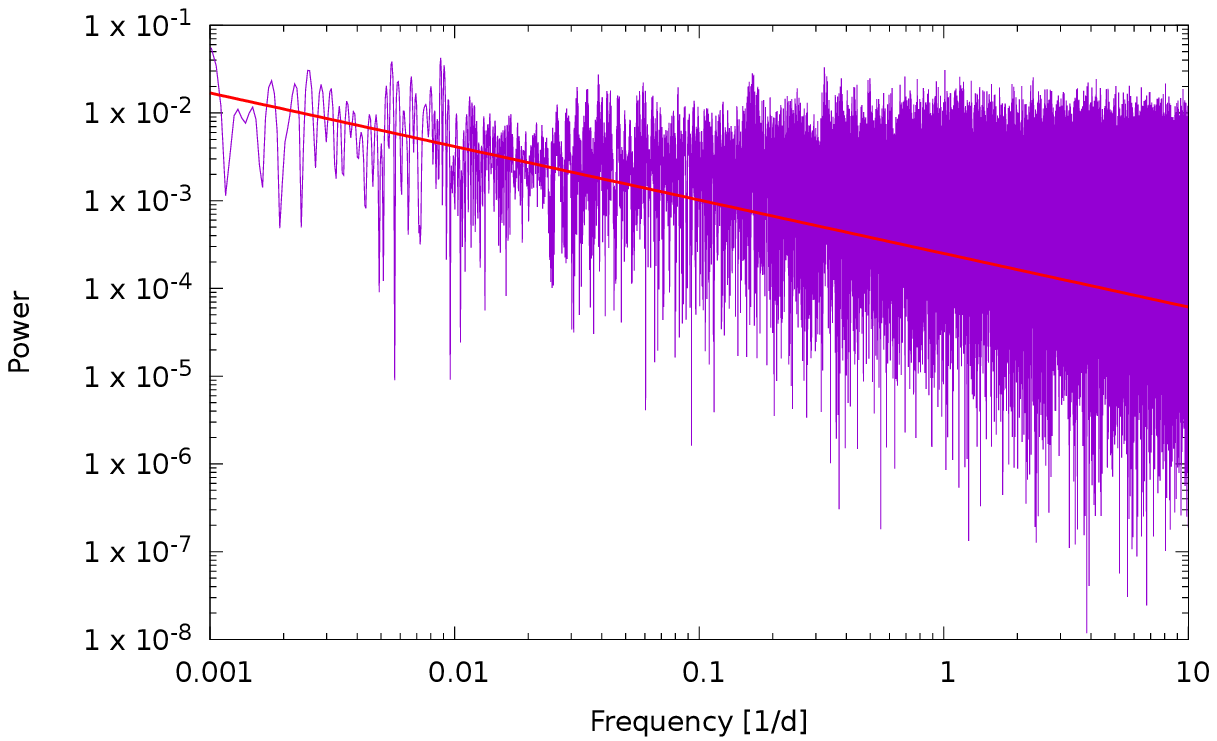}
\vspace{1cm}
\includegraphics[width=6cm,height=4.3cm]{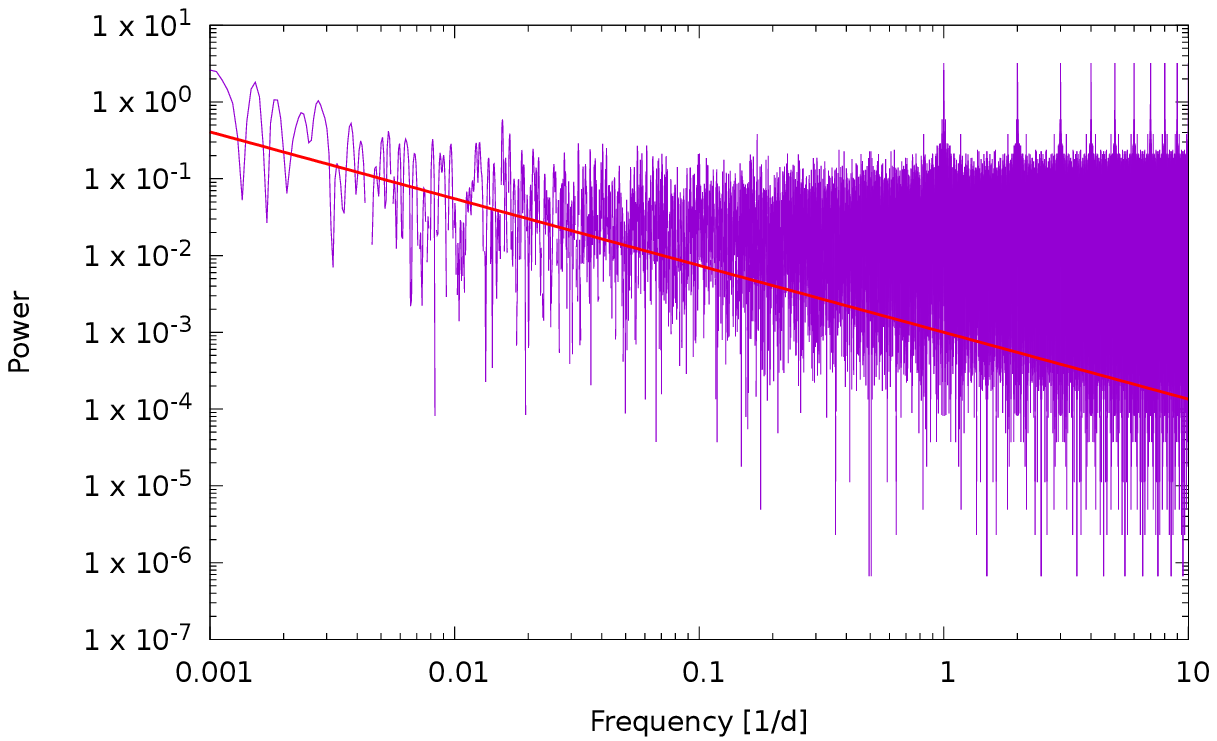}


\caption{ From 
 left to right and top to bottom, the panels in the figure
correspond to radio, optical, X$-$ and \( \gamma \)$-$rays' power spectrum
density (PSD) for the blazar Mrk~501. In all cases, the color of the noise of
the signal is pink, according to the results in Table~\ref{tab:psd-mrk501}.
} \label{fig-psd}

\end{figure}

Table~\ref{tab:psd-mrk501} shows that the application of a PSD and a
DFA to the  multi-frequency light curves of Mrk~501 results in pink
noise.	For completeness, Figure~\ref{fig-psd} shows the PSD plots in
all studied wavelengths. The PSD and DFA analyses were performed using
VARTOOLS, and the fitting of the corresponding \( \alpha \) and \( \beta
\) exponents were calculated using our own C program with the aid of the
GNU Scientific Library (\url{https://www.gnu.org/software/gsl }, {accessed
on 18 January 2024}
) routines.

\begin{table} [H]
\caption{The table shows the obtained values for
the exponents $\alpha$ and \( \beta \) resulting from the PSD and DFA
color of noise analysis to the multi-frequency observations of Mrk~501.
In all cases, the resulting color of noise is pink, according to the
classification presented in Table~\ref{tab:colornoise}, which implies that
the light curves present temporal correlations with random fluctuations.
} \label{tab:psd-mrk501} 
 
 \setlength{\cellWidtha}{\columnwidth/3-2\tabcolsep-0.0in}
\setlength{\cellWidthb}{\columnwidth/3-2\tabcolsep-0.0in}
\setlength{\cellWidthc}{\columnwidth/3-2\tabcolsep-0.0in}

\scalebox{1}[1]{\begin{tabularx}{\columnwidth}{>{\PreserveBackslash\centering}m{\cellWidtha}>{\PreserveBackslash\centering}m{\cellWidthb}>{\PreserveBackslash\centering}m{\cellWidthc}}
\toprule

 \textbf{Band} &
\boldmath{$\alpha$} &  \boldmath{$\beta$}  \\ 
\cmidrule{1-3} 

	Radio &  \(1.24\) $\pm$ \(0.020 \) &  \(1.287 \) $\pm$ \(0.0035\)
\\ Optical &  \(1.16 \) $\pm$ \(0.030 \) & \(0.869 \) $\pm$ \(0.0045\)
\\ X$-$rays & \(0.61 \) $\pm$ \(0.005 \) & \(1.019 \) $\pm$ \(0.0045\) \\
$\gamma$-rays & \(0.87 \) $\pm$ \(0.035 \) & \(0.853 \) $\pm$ \(0.0015\)
\\ 
\bottomrule

 \end{tabularx} }

\end{table}

The power spectrum density and the detrended fluctuation
analysis complement the methods used to search for periodicities. Nevertheless,
the variations in the emission of blazars are more complex. For example,
\citet{Vaughan2016} mention that periodic variations in quasar light
curves have raised the possibility of close binary supermassive black
holes. However, quasars typically display stochastic variability,
making it challenging to identify true periodic candidates. Methods
like Bayesian analysis light curves favors a stochastic process over
sinusoidal variation, though simulations of the quasar PG~1302-102 light
curves demonstrate the occurrence of periodicity in stochastic processes,
which is crucial to calibrate false positive rates when searching for
periodic signals among numerous targets. The studied X$-$ray observations of
PG~1211+143~\citep{Lobban2018} showed time lags where strong correlations
were observed, including complex variability patterns. This phenomenon
was studied as well by~\citet{Vaughan2015}, analyzing X$-$ray variability
in black hole sources including AGN, X$-$ray binaries and UltraLuminous
X$-$ray objects, finding the accretion flow as the source of the observed
X$-$ray variations.







\subsection{Data Fitting via a Jacobi Elliptical Function} \label{jacobi}

\begin{figure} [H]
	   \includegraphics[width=14cm]{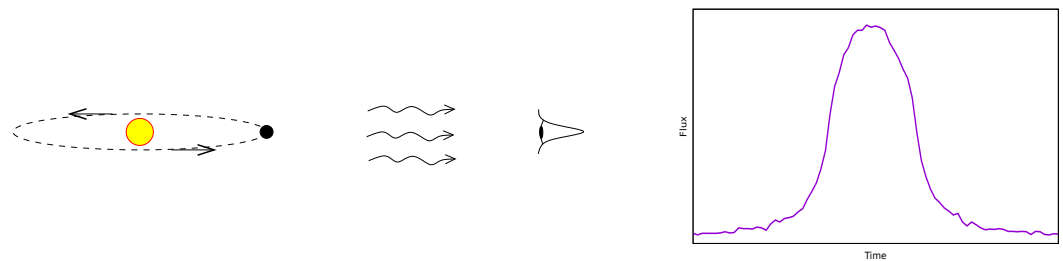}
   \caption{The illustration shows a black hole orbiting
a central spherical source of light that eclipses the radiation
detected by an observer.  For simplicity, and in order to amplify the
magnification effect detected by the observer, the example shown in the
figure has the line of sight of the observer within the plane of orbit.	The right plot shows the radiation flux detected by the observer.
It consists of a numerical simulation of a Schwarzschild black hole
orbiting a fixed, spherical source of light. Over an orbital period,
the passage of the black hole through the line of sight of the observer
magnifies the flux detected. The plotted flux and time are normalized to
numerical units for a spherical source of radius five emitting isotropic
radiation, a Schwarzschild radius of the black hole of one and 
an orbit of radius thirty. The ray-tracing technique used for this
simulation was performed using a squared screen normal to the line of
sight at a distance of one thousand. A video 
of this numerical simulation
can be found at \url{https://archive.org/details/blackhole_magnification}, accessed on 15 March 2024},
and it was produced using a GNU General Public License (GPL) code named
aztekas-shadows, which is under development and will eventually be
available at \url{https://aztekas.org}, accessed on 15 March 2024, copyright \copyright 2020 Gustavo
Magallanes-Guij\'on, Sergio Mendoza and Milton Jair Santiba\~nez-Armenta.

\label{fig-illustration}

\end{figure}

The achromatic periodicity, from radio to \( \gamma \)-rays, of \(
\lesssim\)\(229\, \text{d} \) detected in the previous sections, combined
with the pink noise color, implies that a correlated phenomenon
is producing it. The simplest way in which such a phenomenon may
occur is with an eclipse, as shown in Figure~\ref{fig-illustration},
since an orbiting massive object partially covering the radiation
from the central parts of Mrk~501 will do so in a periodic way.
In what follows, we denote the central supermassive black hole
as the primary black hole and the eclipsing massive object as the
secondary object.  When light is obscured due to a non-relativistic massive
object, it gets dimmed, as happens in a standard eclipse of, say, an exoplanet 
when it passes through the central
star of its planetary system \citep{handbook-exoplanets}.  However,
light is magnified if a relativistic object passes through light beams
produced by a primary source of light (see, e.g., \citep {narayan1996lectures} and references
therein). The following eclipse function, \(e(t)\),
serves as a good empirical way to model an eclipse:

\begin{equation} e(t) = \pm \bigg\{ \Theta(t) - \Theta(t-\pi)  \bigg\}
   \, \mathrm{sn}^2 \left( \frac{ 2 K(m) t }{ \pi } \right)
\label{ecu:jacobi} \end{equation}

\noindent where \( \Theta \) represents the heaviside step function,
\( \mathrm{sn} \) is the elliptic sine or sinus amplitudinis Jacobi
function with module \( m \), such that \( 0 \leq m \leq 1 \),	and \(
K(m) \) is the	complete elliptic integral of the first kind given
via~(see, e.g., \citep {abramowitz-stegun}):

\begin{equation} K(m) = \int_0^{\pi/2}{ \frac{\mathrm{d}\zeta}{\sqrt{1 -
  m \sin^2 \zeta } } }.
\label{complete-elliptic-integral} \end{equation}

\noindent The \( \pm \) sign on the right-hand side of
Equation~\eqref{ecu:jacobi} represents a standard non-relativistic
eclipse, i.e., an eclipse that diminishes the amount of received radiation,
for the minus sign and a magnification relativistic eclipse for the
plus sign.  In the limit that occurs when 
the module \( m \rightarrow 1 \), a single
squared pulse is obtained, and when \( m \rightarrow 0 \) occurs, 
 a sinusoidal
curved profile \mbox{is obtained}.

\begin{figure}[H]


\includegraphics[width=6cm,
  height=4.3cm]{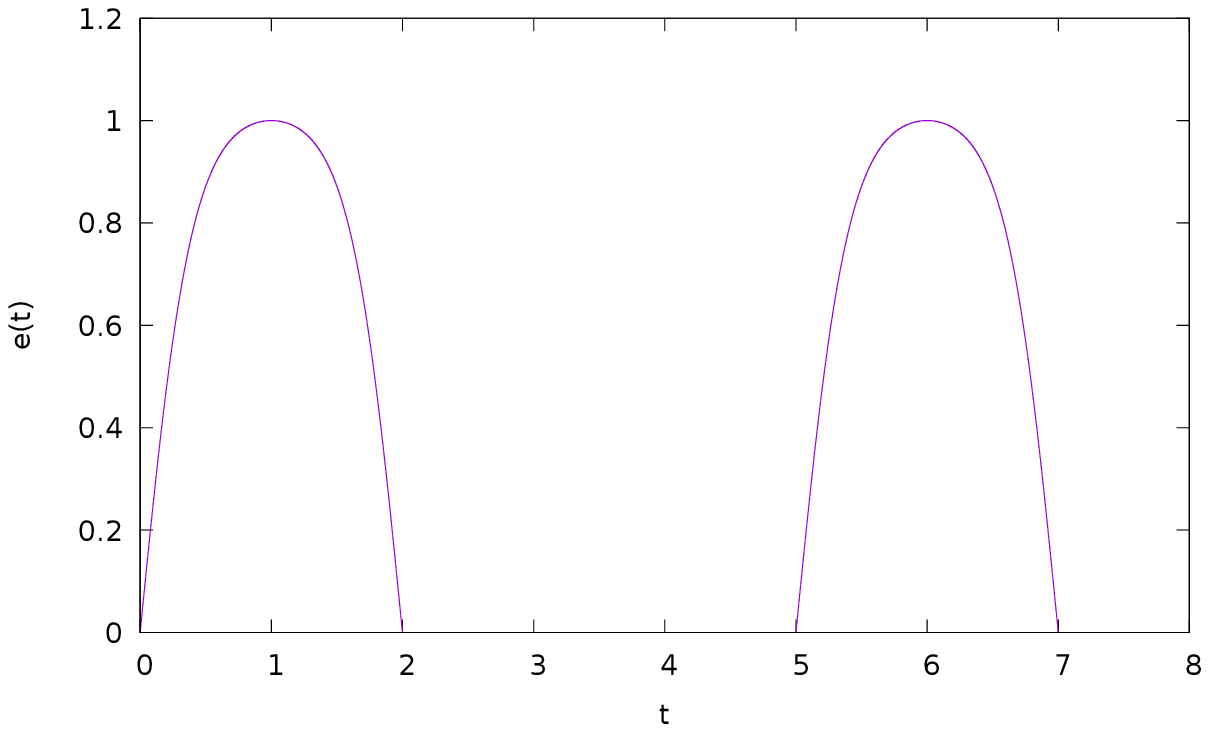}\hspace{1cm}
  \includegraphics[width=6cm, height=4.3cm]{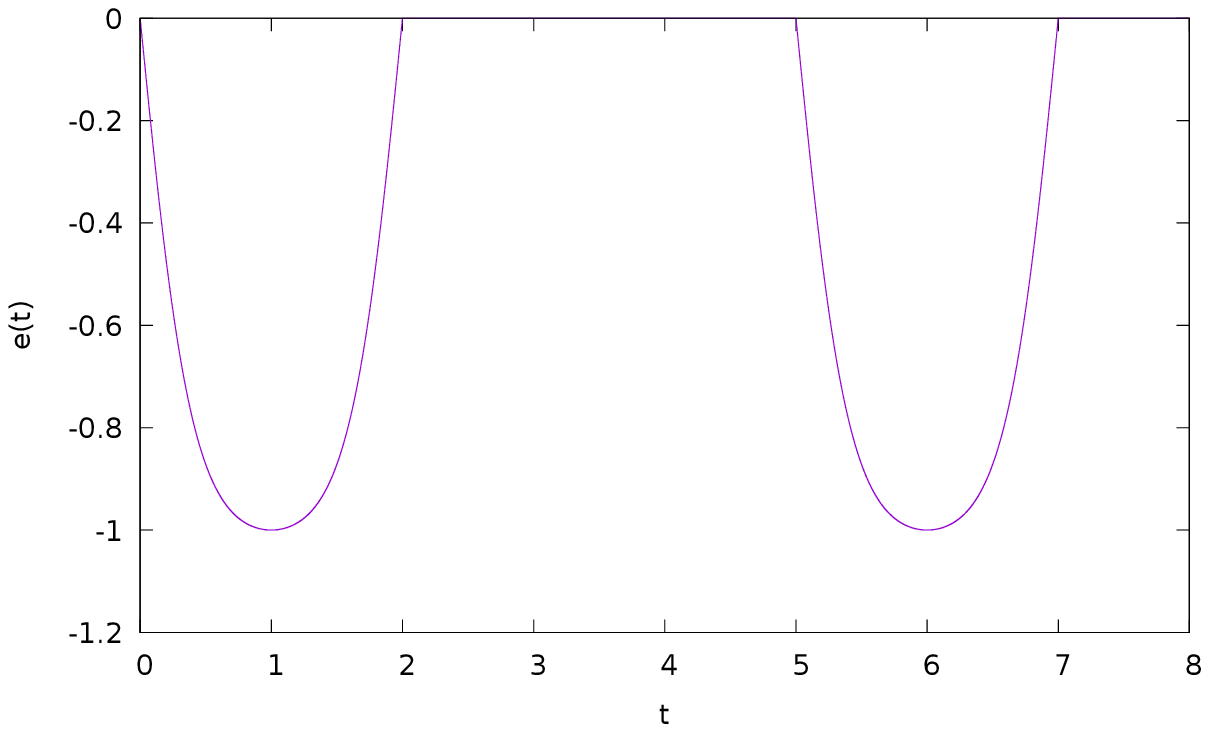}
  \\ \vspace{.5cm} \includegraphics[width=6cm,
  height=4.3cm]{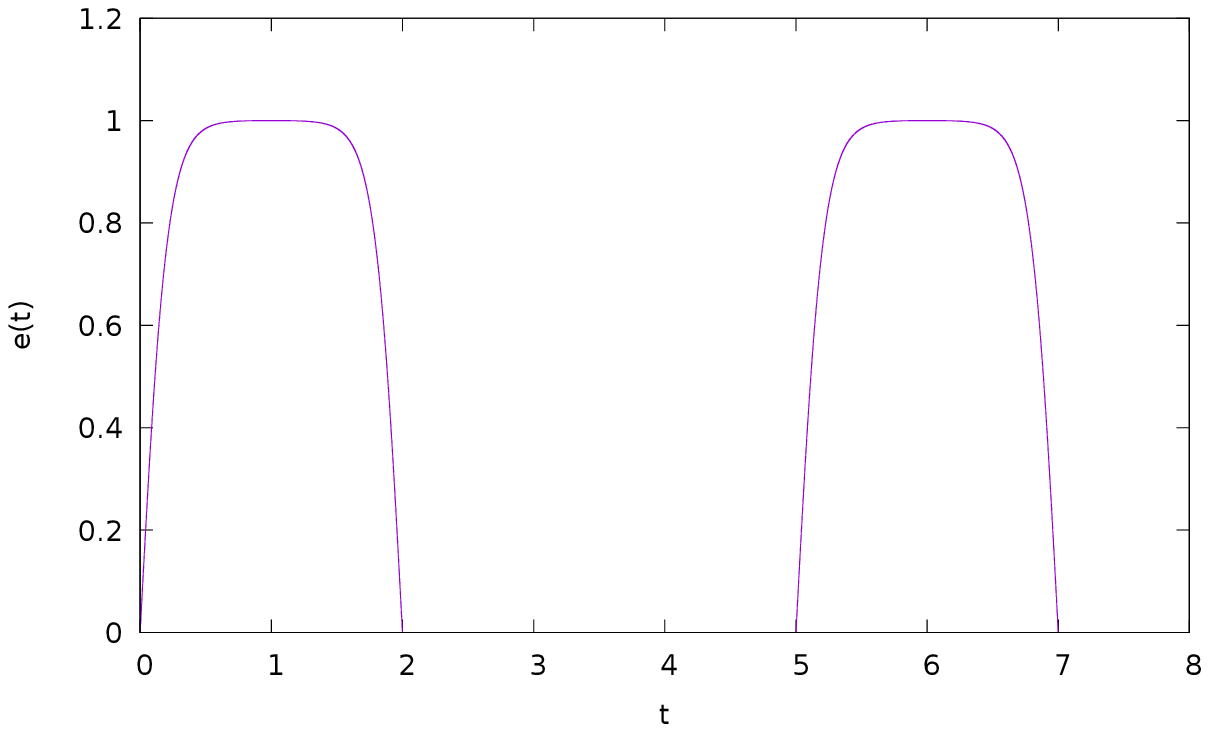}\hspace{1cm}
  \includegraphics[width=6cm, height=4.3cm]{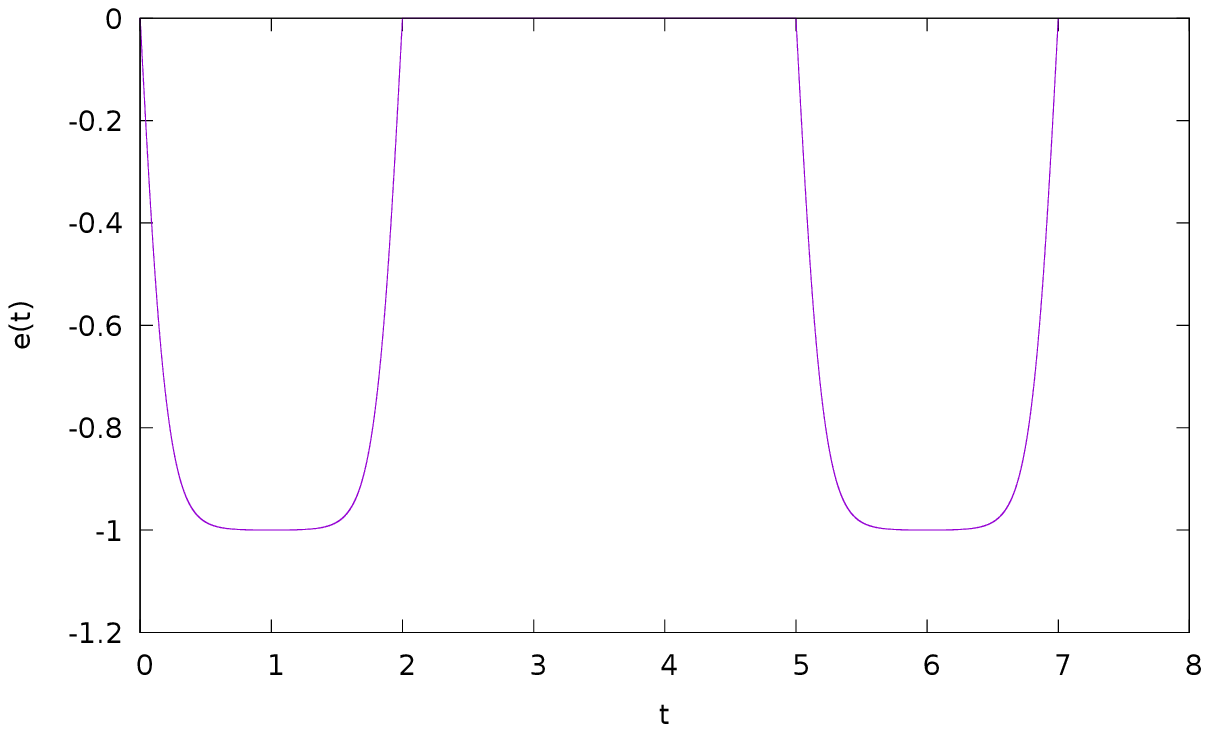}
  \\ \vspace{.5cm} \includegraphics[width=6cm,
  height=4.3cm]{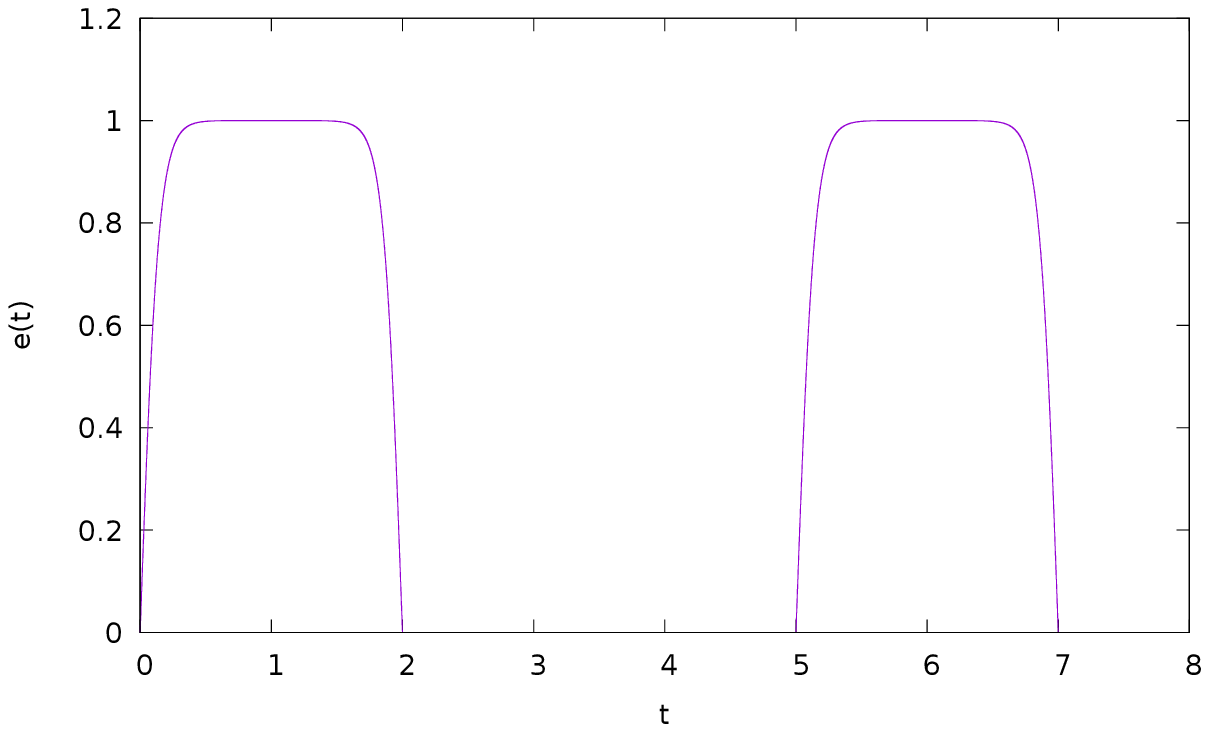}\hspace{1cm}
  \includegraphics[width=6cm,
  height=4.3cm]{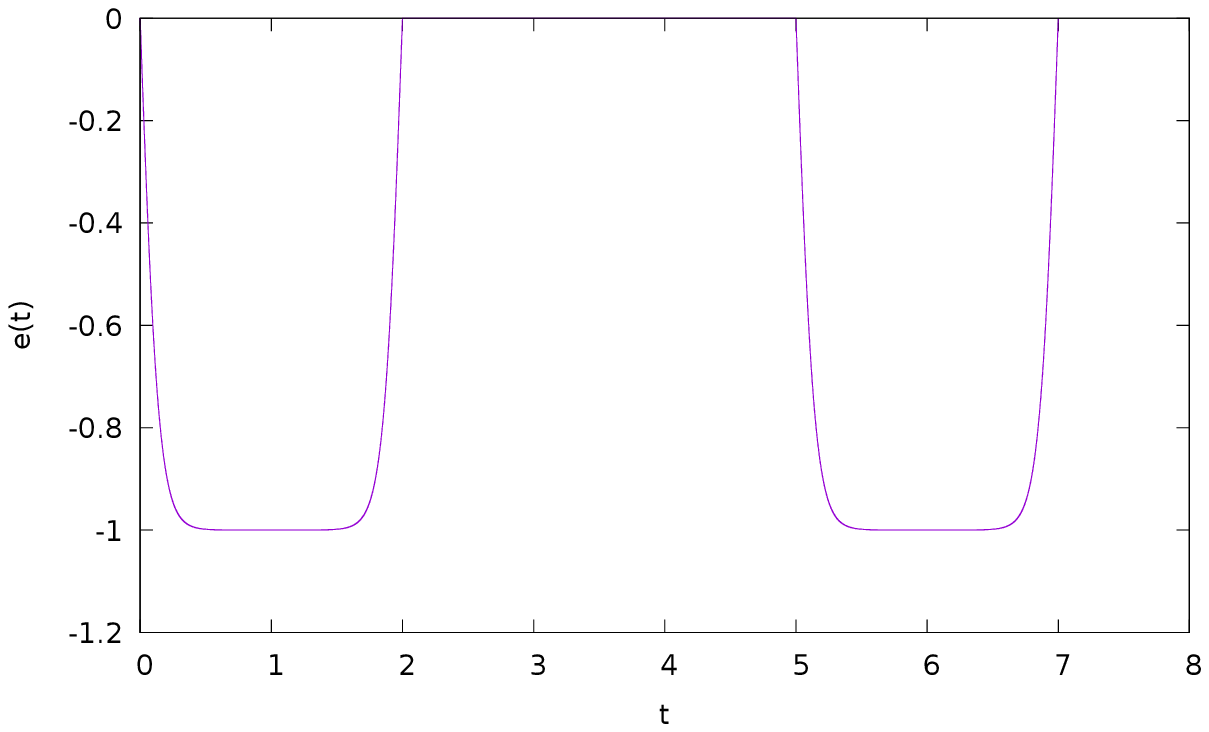} 
\caption{The 
figure shows plots of an artificial eclipse
that occurs two times. The software that produced them is described
in Section~\ref{jacobi} and bears the copyright \copyright 2022 Gustavo
Magallanes-Guij\'on and Sergio Mendoza.	From top to bottom, different
values of the Jacobi elliptic function \( m = 0.9,\, 0.999,\, 0.99999
\) were chosen and for all plots with 
the duration time of the eclipse, \(
t_\text{e} = 2 \), for a quiescent  time, \( t_\text{q} = 3 \), and an
amplitude, \( A = 1 \).	The left column shows relativistic eclipses that
produce magnification, which correspond to a plus sign in the simplified
Equation~\eqref{ecu:jacobi}, and the right column represents a standard,
non-relativistic eclipse, showing the diminishing of the radiation represented
by a minus sign in the \mbox{same equation}.  } \label{jacobi-eclipse-examples}

\end{figure}

To model the obtained periodicity on the light curves of Mrk~501 shown
in the previous sections, we built a C program capable of dealing with
the expectations of an eclipse fingerprint produced on a long-term
light curve based on a modification of Equation~\eqref{ecu:jacobi}.
The program has as free parameters, the amplitude, \( A \), of the eclipse,
the duration time, \( t_\text{e} \), of the eclipse, the quiescent duration
time, \( t_\text{q} \), when the eclipse is not occurring and the number
of occurrences of the eclipse, \( n \).	 We chose the statistical
average \( \langle f \rangle  \) as the zero or quiescent flux of the
light curve.  A set of artificial examples of these eclipses is presented
in Figure~\ref{jacobi-eclipse-examples}.


The data fitting for Mrk~501 was performed using a Monte Carlo
method, based on the program mentioned in the previous paragraph.
Random seeds were used for the unknown parameters of the program, using
\( 10^6 \) attempts for each light curve for Gaussian noise.	Table
\ref{montecarlo-mrk501} shows the results of the fit for each waveband.


\begin{table} [H]

\caption{The table shows the best fit of the eclipse model using a Monte
Carlo method for the multiwavelength data of Mrk~501. The columns represent
electromagnetic wavebands, the eclipse amplitude, \(A \), the elliptic Jacobi
function module, \( m \), the duration time of the eclipse, \( t_\text{e} \),
the interval of time where the eclipse is not occurring, i.e., the quiescent
time, \( t_\text{q} \), the number of times, \( n \), the eclipse occurred,
the total periodicity, \( p := t_\text{e} + t_\text{q} \), the mean value of
the flux, \( \langle f \rangle  \), and the dimensionless brightness
magnification produced via the eclipse  \( A / \langle f \rangle  \).
\label{montecarlo-mrk501} } 


\begin{adjustwidth}{-\extralength}{0cm}
\setlength{\cellWidtha}{\fulllength/9-2\tabcolsep-0.2in}
\setlength{\cellWidthb}{\fulllength/9-2\tabcolsep-0.0in}
\setlength{\cellWidthc}{\fulllength/9-2\tabcolsep-0.0in}
\setlength{\cellWidthd}{\fulllength/9-2\tabcolsep-0.0in}
\setlength{\cellWidthe}{\fulllength/9-2\tabcolsep+0.1in}
\setlength{\cellWidthf}{\fulllength/9-2\tabcolsep-0.0in}
\setlength{\cellWidthg}{\fulllength/9-2\tabcolsep+0.1in}
\setlength{\cellWidthh}{\fulllength/9-2\tabcolsep-0.0in}
\setlength{\cellWidthi}{\fulllength/9-2\tabcolsep-0.0in}

\scalebox{1}[1]{\begin{tabularx}{\fulllength}{>{\PreserveBackslash\centering}m{\cellWidtha}>{\PreserveBackslash\centering}m{\cellWidthb}>{\PreserveBackslash\centering}m{\cellWidthc}>{\PreserveBackslash\centering}m{\cellWidthd}>{\PreserveBackslash\centering}m{\cellWidthe}>{\PreserveBackslash\centering}m{\cellWidthf}>{\PreserveBackslash\centering}m{\cellWidthg}>{\PreserveBackslash\centering}m{\cellWidthh}>{\PreserveBackslash\centering}m{\cellWidthi}}
\toprule

 \textbf{Band} &
\boldmath{\(A \)}	&\boldmath{ \( m \) }& \boldmath{\( t_\text{e} \)} &	\boldmath{\( t_\text{q} \) }& \boldmath{\( n \)} & \boldmath{\(
p \) }& \boldmath{\( \langle f \rangle  \)} & \boldmath{\( A / \langle f \rangle \)} 
\\ & &	&
\textbf{Days} &	\textbf{Days} &	 & \textbf{Days}  &  & \\ 
\cmidrule{1-9} 

Radio & $0.09 \pm 0.04$ & $0.85
\pm 0.2 $ & $88.33 \pm 2.83$ & \mbox{130.20 $\pm$ 12.77} &   $16.43 \pm 1.80$
& $218.53 \pm 15.6 $ & $1.11 \pm 0.022$ & $0.08 \pm 0.04$ \\ Optical  &
$1.18 \pm 0.02 $ & $0.14 \pm 0.21$ & $89.76 \pm 3.60$ & $133.18 \pm 0.09$
& $35.03 \pm 0.43$ & $222.94 \pm 3.69$ & $13.48 \pm 1.34$ & $0.08 \pm
0.01$ \\ X$-$rays & $0.99 \pm 0.44$ & $0.82 \pm 0.02$ & $89.37 \pm 0.27$
& $137.05 \pm 3.61$ &  $19.96 \pm 0.51$ & $226.43 \pm 0.41$ & $2.46 \pm
0.68$ & $0.40 \pm 0.29$\\ \(\gamma\)-rays  & $0.99 \pm 0.07 $ & $0.20
\pm 0.17$ & $86.25 \pm 0.32$ & $142.10 \pm 4.45$ & $17.69 \pm 0.41 $ &
$228.36 \pm 4.77$ & $0.13 \pm 0.48$ & $7.30 \pm 26.65$ \\

\bottomrule

\end{tabularx} }

\end{adjustwidth}

\end{table}


\section{Discussion}  \label{discussion}

A different analysis of the long-term multiwavelength (from radio
to \(\gamma\)-rays) light curves of Mrk~501 show a common achromatic
periodicity of \( \lesssim\)\(229\, \text{d} \). The Monte Carlo fitting
technique, assuming a relativistic eclipse as the cause of this
periodicity, shows a periodicity of \( \sim\)\(224 \, \text{d} \). The
eclipse is produced by a secondary supermassive black hole orbiting 
the primary supermassive black hole, i.e., about the central engine
of Mrk~501. The reasoning for this conclusion is summarized in the
following paragraphs.

The periodicities found with the RobPer and L-S algorithms described
in Section~\ref{periodograms} are all consistent in all frequencies of
Mrk~501 with average values in radio, optical, X$-$rays and \(\gamma\)-rays,
respectively, given via	\( 228.03 \), \(226.77 \), \( 223.20 \) and  \(
238.90 \) days, according to the results of Table~\ref{tab:per}. The mean
value for these periodicities is \(229.225 \,\text{d}\).

With the use of the  VARTOOLS software described in
Section~\ref{vartools}, we found that, for the AoV, AoV-h, BLS and DFT
routines, the periodicity value lies in the intervals 226.73--229.2,
219.7--228, 220.960--240.404 and 222.374--242.818 days, respectively,
according to the results presented in Figures~\ref{fig-aov}
and~\ref{fig-bsl-dft}.

The fact that the color of the signal noise in the light curves of
Mrk~501 presented in Section~\ref{power-spectrum} is pink means that,
for each particular waveband, there is a robust oscillation (Brownian
noise color) with a periodicity accompanied by a random signal (white
\mbox{noise color)}.

\begin{figure} [H]
  \includegraphics[width=13.5cm]{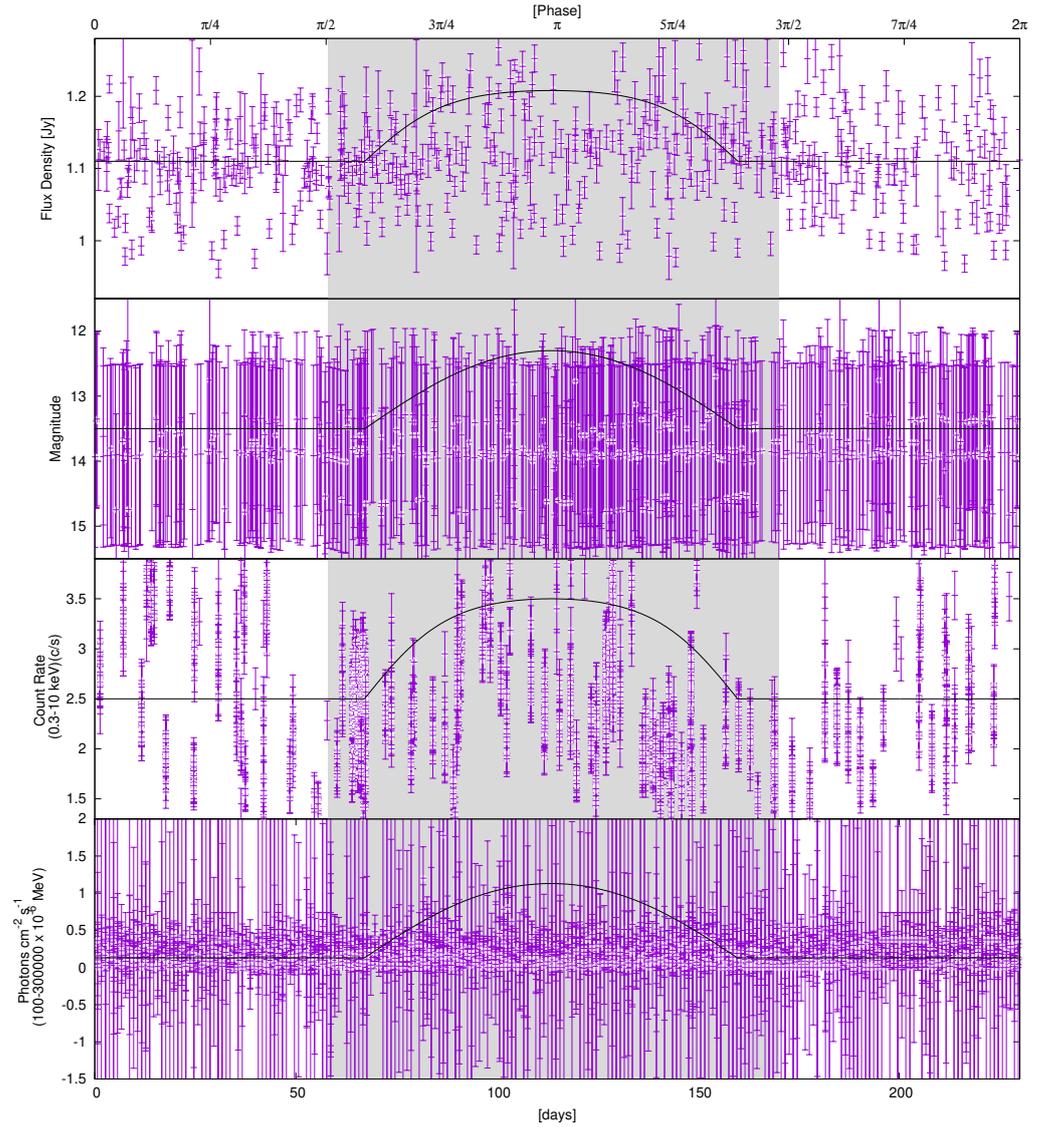} 
  \caption{From 
 top to bottom, the figure shows \(224
\, \text{d} \) folded light curves for radio, optical, X$-$ and
$\gamma$$-$rays. The solid curves are the best fit eclipse model
described in Section~\ref{jacobi} constructed using the results of
Table~\ref{montecarlo-mrk501}.	The shaded zone in each panel represents
the duration of the eclipse.  The dotted horizontal lines represent the \(
1\sigma \) significance level.	}

\label{fig-all-freq}

\end{figure}

Due to the achromatic nature of the found periodicity of \( \lesssim\)\(229 \, \text{d} \), we modeled this periodicity as a relativistic
eclipse caused by an orbiting supermassive black hole about the central
engine of Mrk~501. The results of Table~\ref{montecarlo-mrk501} show
that this model is quite coherent in the fitting of the long-term
multi-frequency light curves. The only small inconsistency found
is with the dimensionless brightness magnification \( A / \langle f
\rangle \) presented in X$-$rays and more prominent in \( \gamma \)-rays.
This is most probably due to the large errors reported in \( \gamma
\)-rays and the large gaps that appear in the X$-$ray light curve.
Figure~\ref{fig-all-freq} shows a time-folding of the multi-frequency
light curves with the corresponding eclipse function using the results
of Table~\ref{montecarlo-mrk501}. The shaded region represents the time
duration, \( t_\text{e} \), of the eclipse. The dotted horizontal lines
represent a \(1\sigma\) significance level. It is important to note that
the radio, optical and X$-$ray panels in Figure~\ref{fig-all-freq} do not
contain all the used data points at a \( 3\sigma  \) confidence level.
In other words, they represent zooming in to the light curves in order to
emphasize the detected eclipse. The large error bars in the \( \gamma
\)-ray make the bump on the eclipse function cross the \( 3\sigma \)
confidence level.

As mentioned in the introduction, shockwave interactions and
motions inside the jet could cause periodic oscillations on the
light curves, and although the periodicity in this case may appear,
in principle, as an achromatic effect, it is extremely unlikely that
this is the case for Mrk~501 since the reported amplitudes, \( A \), in
Table~\ref{montecarlo-mrk501} are quite similar to one another, a property
unlikely to occur in periodic shock wave formation and interactions due
to the complexity of the radiation produced at each particular frequency.


In this article, we found an achromatic periodicity of \( \sim \)224~d in
the radio, optical, X$-$ and \( \gamma \)$-$rays light curves of Mrk~501,
which we modeled as an eclipsing event produced by a massive (secondary)
supermassive black hole orbiting the (primary) supermassive  black
hole of  Mrk~501. Since the eclipse is produced by a relativistic
object (the secondary black hole), the radiation brightness is magnified 
(contrary to what occurs in, e.g., a solar eclipse on Earth).

Using the results of our eclipse model in Table~\ref{montecarlo-mrk501}
and Kepler's third law for a circular orbit given via

\begin{equation} 
v = \frac{ r_\text{orbit} }{ p }  = \sqrt{ \frac{
  G M_\text{prim} }{ r_\text{orbit}
  } }, 
\label{kepler}
\end{equation}

\noindent where v is the velocity of the orbiting test mass (secondary
black hole), \( r_\text{orbit} \) and \( p \) are the radius and period
of the orbit, \( M_\text{prim} \) is the mass of the central supermassive
black hole and \( G \) represents Newton's constant of gravitation,
the following conclusions can be drawn about the results obtained in
this article:

\begin{itemize} \item The mass of the central (primary) black hole in
  Mrk~501 is
	\( M_\text{prim}\sim\)\(10^9 \, \text{M}_\odot \) \citep{rieger03},
	which means that its gravitational radius is 
	 \( r_\text{g-prim}\sim\)\(20 \, \text{au}\).
  \item The radius of the orbit of the eclipsing (secondary) binary
	black hole is \( r_\text{orbit} \sim\)\(200 \, \text{au} \sim10 \,
	r_\text{g-prim} \). 
	
	Using a full relativistic approach for
	Schwarzschild's space--time, Equation~\eqref{kepler} can be written
	as follows~\citep{mendoza-gravitation}:
	\begin{equation}
	  v^2 = \frac{ 1 }{ 2 } c^2 \left( r_\text{g-prim} / r_\text{orbit}
	  \right) \left( 1 - 3 r_\text{g-prim} 
	    / 2 r_\text{orbit} \right), 
	\label{kepler-relativistic}
	\end{equation}
	\noindent The above equation is cubic for the radius
	\( r_\text{orbit} \), for which the only real solution is given via the following:
	\begin{equation}
	  r_\text{g-prim} / r = \left( A \sqrt{ 1 + A } \right)^{1/3} - A /  
	    \left( A \sqrt{ 1 + A } \right)^{1/3},
	\label{real-root}
	\end{equation}
	\noindent where
	\begin{equation}
	   A := r_\text{g-prim}^2 / c^2  p^2.
	\end{equation}
        By using the same input values as the ones we used for the
	non-relativistic Kepler formula, we obtain \( r_\text{orbit}
	\sim\)\(5 r_\text{g-prim} \), which is of the same order of magnitude as the
	\( 10 r_\text{g-prim} \) that we obtained with the simple Kepler
	relation.
	
  \item The orbital period of the eclipsing binary black hole is \(
	\sim\)\(224  \, \text{d} \).
  \item The orbital velocity of the eclipsing binary black hole is \(
	\sim\)\(0.3 \% \) of the speed of light, i.e., \( \sim\)\(3 \times 10^6
	\, \text{km} / \text{h} \).
  \item The brightness magnification of the radiation produced by an
	eclipse due to the secondary black hole is \( \gtrsim\)\(10 \% \).
  \item The coalescence time of the binary system is approximately
	given via the following ~\citep{taylor}: 
	 \begin{equation}
	   t_\text{coal} \approx \frac{ 1 }{ 50 } \frac{ c^5 }{G^3}
	   \frac{ r_\text{orbit} }{ M_\text{prim}^2 M_\text{sec} } \sim
	   \frac{ 10^8 \text{yrs} }{  \left( M_\text{sec} / M_\odot
	   \right) },
	 \label{coallescence} \end{equation} \noindent where \(
	 M_\text{sec} \, \text{(}\ll M_\text{prim}\text{)}  \) is the mass
	 of the secondary or eclipsing black hole. Since the periodicity
	 found corresponds to a few decades of observation---meaning that
	 the orbit is sufficiently stable---it is reasonable to assume
	 that \( t_\text{coal} \gtrsim 10^3 \), which means that the
	 mass of the secondary black hole \( M_\text{sec} \lesssim 10^5
	 \text{M}_\odot \).
   \item The results of
~\eqref{kepler-relativistic} imply that the
         orbit of the secondary black hole is half the value obtained from
	 the Newtonian calculation, and so its orbital frequency is about \(
	 2^{ 3 / 2 } = 2.83\)\(\sim\)\(3 \) its Newtonian value.  
	 Since the frequency, \( f \), of the of the binary's quadrupolar
	 gravitational waves is about a factor of \( 2 \) larger than the orbital
	 frequency~(see, e.g., \citep{gravitational-wave}), then
	 \begin{equation}
	   f \sim  6 \sqrt{ G M / r_\text{orbit}^3 } \, , 
         \end{equation}
	 \noindent where the total mass of the binary system is
	 given via \( M = M_\text{prim} + M_\text{sec}
	 \approx M_\text{prim} \). In other words,
	 \begin{equation}
	    f \sim 6 \times 10^{5} \left( \frac{ M_\odot }{
	    M_\text{prim} } \right) \, \text{Hz},
	 \label{frequency-bh-numbers} 
	 \end{equation} 
	 \noindent which, for a value of \( M_\text{prim} 
	 \)\(\sim\)\(10^9 M_\odot \), yields
	 \( f\)\(\sim\)\(6 \times 10^{-4} \, \text{Hz} \). This suggests that
	 the frequency of the emitted gravitational waves is just
	 above \( 0.1 \, \text{mHz}\), which is within the range of the
	 forthcoming LISA space-based interferomenter, and as such, the
	 prediction of a binary black hole system in Mrk~501 should be
	 considered a science target for future LISA observations.
\end{itemize}

The magnification \( \mu \) of the radiation is the ratio of the
Einstein radius, \( \theta_\text{E} \), to the distance, \( \beta \), from
the line of sight connecting the observer to the lens~\citep{dodelson}.
Since the secondary eclipsing black hole mass \(\sim\)\(10^5 M_\odot
\), then \( \theta_\text{E}\)\(\sim\)\(10^{-8} \), and so, in order to get a
magnification of \( \mu = 1.1 \), then \( \beta \gtrapprox 10^{-8} \).
In other words, it would seem that the primary and secondary black
holes should be quite well aligned with our line of sight. However,
since Mrk~501 is a blazar, we are observing the source within an
angle of no more than \( 30^\circ \) from the emitted jet (in fact, for
Mrk~501, the observing angle is between \( 15^\circ \) and \( 25^\circ \),
as reported by~\citet{angle}), and so the radiation would be a combination
of the radiation processes occurring close to the central engine plus
the relativistic jet. This extended emission from the relativistic jet
means that the chances of having an 10\% amplification increase
.

\vspace{12pt}
\authorcontributions{~~}
Conceptualization, G.M.-G. and S.M.; methodology, G.M.-G. and S.M..;
software, G.M.-G.; validation, G.M.-G.. and S.M.; formal analysis, G.M.-G.
and S.M.; investigation, G.M.-G. and S.M.; resources, G.M-G. and S.M.; data
curation, G.M.-G. and S.M.; writing---original draft preparation, G.M.-G.
and S.M.; writing---review and editing, G.M.-G. and S.M.; visualization,
G.M.-G. and S.M.; supervision, G.M.-G. and S.M.; project administration,
G.M.-G. and S.M.; funding acquisition, S.M. All authors have read and agreed to the published version of the manuscript.

\funding{~~}
This research was funded by PAPIIT DGAPA-UNAM grant number IN110522 and
CONAHCyT grant numbers 378460 and 26344.

\dataavailability{~~} 

\acknowledgments{ This work was supported by 
The authors thank Erika Benitez for discussions and comments while
preparing this work. We are also grateful to Milton Santiba\~nez-Armenta
for discussions and programming help with the black hole eclipsing
video that supports the results of this work.  We thank the OVRO
40-m monitoring program~\citep{Richards2011ApJ} for the radio database used. The
public OVRO database is supported with private funding from the California
Institute of Technology and the Max Planck Institute for Radio Astronomy,
as well as NASA grants NNX08AW31G, NNX11A043G and NNX14AQ89G and NSF grants
AST-0808050 and AST-1109911.  We also thank the variable star database
of observations from the AAVSO International Database, which has been contributed to by
observers worldwide. We are grateful for the public data observations from the
Swift data archive and the \textit{Fermi} Gamma-Rays Space Telescope
collaboration for the public database used in this work.}

\conflictsofinterest{~~} 
The authors declare no conflict of interest.

\begin{adjustwidth}{-\extralength}{0cm}

\reftitle{References}

\PublishersNote{}
\end{adjustwidth}

\end{document}